\pdfoutput=1

\documentclass[preprint]{revtex4-2}
\usepackage{graphicx}
\setcitestyle{round}

\usepackage{color}
\usepackage{bm}
\usepackage{braket}
\usepackage{amsmath}

\usepackage{natbib}

\usepackage{caption}
\def\cblue{\color{blue}}

\begin{document}

%\preprint{}

\title{Colossal anomalous Nernst effect in a correlated noncentrosymmetric kagome ferromagnet}
\author{T. Asaba$^{1*}$, V. Ivanov$^{2}$, S. M. Thomas$^{1}$, S.Y. Savrasov$^{2}$, J. D. Thompson$^{1}$, E. D. Bauer$^1$, F. Ronning$^{3*}$}

\affiliation{
$^1$Materials Physics and Applications Division, Los Alamos National Laboratory, NM, 87545 USA \\
$^{2}$Department of Physics, University of California, Davis, CA 95616, USA \\
$^{3}$Institute for Materials Science, Los Alamos National Laboratory, NM, 87545 USA \\
}
\date{\today}
\begin{abstract}
Analogous to the Hall effect, the Nernst effect is the generation of a transverse voltage due to a temperature gradient in the presence of a perpendicular magnetic field. The Nernst effect has promise for thermoelectric applications and as a probe of electronic structure. In magnetic materials, a so-called  anomalous Nernst effect (ANE) is possible in zero magnetic field. Here we report a colossal ANE reaching 23 $\mu$V/K in the ferromagnetic metal UCo$_{0.8}$Ru$_{0.2}$Al. Uranium's $5f$ electrons provide strong electronic correlations that lead to narrow bands, which are a known route to producing a large thermoelectric response. Additionally, the large nuclear charge of uranium generates strong spin-orbit coupling, which produces an intrinsic transverse response in this material due to the Berry curvature associated with the relativistic electronic structure. Theoretical calculations show that at least 148 Weyl nodes and two nodal lines exist within $\pm$ 60 meV of the Fermi level in UCo$_{0.8}$Ru$_{0.2}$Al. This work demonstrates that magnetic actinide materials can host strong Nernst and Hall responses due to their combined correlated and topological nature.
\end{abstract}

\maketitle

{\bf Introduction} 

Because electrons carry both heat and charge, the transport of one is necessarily related to the other. Thermoelectric applications aim to exploit this fact by converting heat or charge flow into electric voltages or temperature gradients \cite{bell2008cooling,he2017advances}. In the presence of a magnetic field transverse responses are possible. It has been argued that the transverse geometry allows one to convert heat into electrical energy much more efficiently than the conventional longitudinal geometry \cite{sakai2018giant,sakuraba2013anomalous,bell2008cooling}. In conventional materials, however, the amplitude of the transverse anomalous Nernst effect (ANE) has been too small to realize practical thermoelectric devices.

Recently, it has been recognized that the Berry curvature associated with the Bloch waves of electrons generates an additional and potentially very large transverse response in magnetic materials \cite{nagaosa2010anomalous, nakatsuji2015large,nayak2016large,li2017anomalous,ikhlas2017large}. This is exemplified by the full-Heusler ferromagnet Co$_2$MnGa whose ANE at room temperature is 6 $\mu$V/K and reaches to a record 8 $\mu$V/K at $T$ = 400 K \cite{sakai2018giant}. This value is larger than typical ferromagnets, and theoretical calculations reveal that the large ANE is a consequence of the intrinsic Berry curvature contribution. After the triumph of Co$_2$MnGa, a natural question arises: Can one further enhance or maximize the size of the ANE for applications or due to fundamental constraints? In this paper, we examine a uranium compound, which has multiple favorable attributes for enhancing the Berry curvature and ANE, and indeed we find an ANE, which is even four times larger than found in Co$_2$MnGa.

The Nernst Effect is related to the thermoelectric ($\bar{\alpha}$) and conductivity ($\bar{\sigma}$) tensors through the expression $S_{xy} = {\cblue -}E_{y}/\nabla_{x}T = (\sigma_{xx}\alpha_{xy} - \sigma_{xy}\alpha_{xx})/(\sigma_{xx}^2 + \sigma_{xy}^2)$. The transverse contribution to the thermoelectric and conductivity tensors is directly connected to the Berry curvature $\Omega_B$ through

\begin{equation} \label{AHC}
{\sigma_{xy}} = -\frac{e^2}{\hbar}\int_{BZ}\frac{d^3k}{(2\pi)^3}f(k)\Omega_B(k)
\end{equation}
\begin{equation} \label{APC}
\alpha_{xy} = \frac{ek_B}{\hbar}\int_{BZ}\frac{d^3k}{(2\pi)^3}s(k)\Omega_B(k)
\end{equation}
where $k_B$ is Boltzmann's constant, $e$ is the charge of an electron, $\hbar$ is Planck's constant $h$ divided by 2$\pi$, $f(k)$ is the Fermi-Dirac distribution function, $s(k)=-f(k)ln(f(k))-(1-f(k))ln(1-f(k))$, and $\int_{BZ}$ is the integration over the Brillouin zone. At low temperatures, these two quantities are related through the Mott formula \cite{cutler1969observation,xiao2006berry}.

An examination of these equations reveals that large Berry curvatures and small Fermi energies $E_F$ will generate large Hall and Nernst effects, in general. As elaborated on below, to enhance the Berry curvature contribution to the ANE and AHE we therefore look for a system with the following properties: (1) a large spin-orbit coupling,  (2) strong electronic correlations, (3) ferromagnetic order, and (4) a kagome lattice structure.

The Berry curvature depends strongly on the spin-orbit coupling (SOC). Increasing the strength of the SOC increases the potential occurrence of band inversions, which generate regions of potentially diverging Berry curvature in the Brillouin zone. Depending on the symmetry of the inverted bands, nodal lines and/or Dirac or Weyl nodes may appear. Generally, the strength of the SOC increases with increasing atomic number, and hence we look for materials near the bottom of the periodic table. 

To maximize the effects of larger SOC requires multiple bands to reside in the vicinity of the Fermi energy. This can be accomplished through strong electronic correlations, which will shift multiple band crossing points closer to the Fermi energy, thereby enhancing the Berry curvature. Furthermore, the renormalized band structures due to electronic correlations lead to smaller effective Fermi energies, which is known to promote large Seebeck and Nernst responses \cite{behnia2004thermoelectricity,behnia2016nernst}. The need for both strong correlations and large SOC leads us naturally to heavy fermion systems composed of $4f$ and $5f$ elements.  

Without broken time-reversal symmetry a transverse Hall or Nernst response is identically zero. Hence, in the absence of an applied magnetic field, magnetic order is required for a finite ANE or AHE. Although possible in a system with antiferromagnetic order \cite{li2017anomalous,ikhlas2017large,wuttke2019berry,xu2020finite}, the requisite absence of symmetries is guaranteed in a ferromagnetic system.

In a system with antiferromagnetic interactions, magnetic frustration present in a kagome lattice can lead to additional band renormalization. Though magnetic frustration is presumably absent in a ferromagnetic material, the unrenormalized hopping on a kagome lattice coincidentally produces one flat band. Perhaps this explains in part, why large AHE and ANE responses have been recently observed in a number of kagome compounds \cite{liu2018giant,kida2011giant,ye2018massive,nakatsuji2015large,xu2015intrinsic}.

%%%%%%%%%Figure1$$$$$$$$$$$
\begin{figure}[p]
	\includegraphics[width=\linewidth]{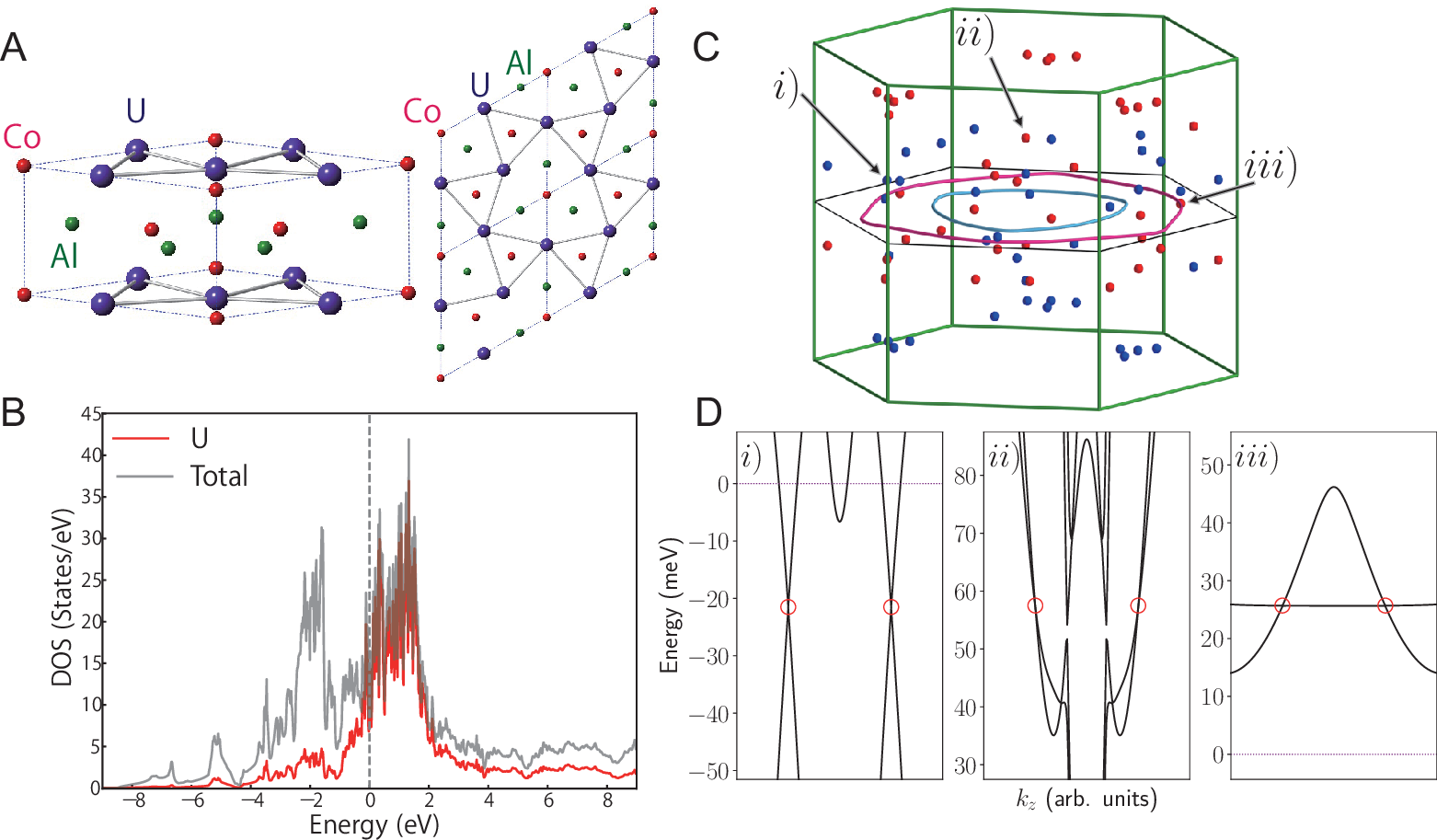}
	\caption{\label{figbasic} 
			{\textbf{Crystal structure and theoretical calculations of UCo$_{0.8}$Ru$_{0.2}$Al.} (A) Crystal structure of UCo$_{1-x}$Ru$_{x}$Al. (left) The unit cell of UCoAl viewed from the [111] direction. A Co atom is replaced by a Ru atom upon doping. (right) The top view of UCoAl. The solid bonds indicate the tilted kagome layer formed by uranium atoms. (B) Density of states (DOS) including partial DOS for UCo$_{0.8}$Ru$_{0.2}$Al. (C) Nodal lines and Weyl points  within $\pm$ 60 meV of the Fermi energy in UCo$_{0.8}$Ru$_{0.2}$Al. For clarity, Weyl pairs with a $k_z$ separation $<$ 0.2 $2\pi/c$ are not shown. (D) Band plots for examples of i) Type-I, ii) Type-II and iii) critically tilted Weyl points (red circles), identified with arrows in (C), plotted on the interval $-2k_W < k_z < 2k_W$, where $2k_W$ is the Weyl pair separation along the $k_z$ direction.}
}
\end{figure}
%%%%%%%%%Figure1$$$$$$$$$$$

In our exploration of materials we found that the layered distorted kagome lattice system UCo$_{1-x}$Ru$_{x}$Al in the ZrNiAl structure type (shown in Fig. \ref{figbasic}A) satisfies all the above criteria. Additionally, the crystal structure is non-centrosymmetric, which can generate pairs of Weyl nodes with opposite chiralities. By themselves such Weyl nodes do not contribute to the AHE and ANE, but they can be further split by the magnetic order. Both parent compounds UCoAl and URuAl are paramagnetic, but much of their solid solution UCo$_{1-x}$Ru$_{x}$Al is ferromagnetic \cite{andreev1997ferromagnetism,pospivsil2016properties}. Members of the entire alloy series show relatively large Sommerfeld coefficients $\gamma$ of $\sim$50 mJ/mol K$^2$ \cite{veenhuizen1988magnetic,matsuda2000transport,pospivsil2016properties,SI}, indicating that charge carriers are moderately heavy.  Magnetic X-ray and photoemission measurements also certify the presence of electronic correlations \cite{antonov2003electronic,fujimori2016electronic}. Furthermore, previous work has shown significant anomalous Hall conductivity (AHC) for dopings close to the quantum-phase transition \cite{pospivsil2016properties}. Here we examine x=0.2 with a $T_C$ of 56 K, close to the optimal value in the series. 

{\bf Results} 

Our spin-polarized theoretical calculations including spin-orbit coupling support our initial conjecture. A partial density of states of UCo$_{0.8}$Ru$_{0.2}$Al shown in Fig. \ref{figbasic}B reveals a narrow band of uranium $f$-states dominating the electronic structure near $E_F$. Electronic correlations are expected to further reduce this bandwidth \cite{SI}. The  magnitude of the calculated anomalous Hall conductivity due to the Berry curvature is found to be $\sim$ 2,000 $\Omega^{-1}$cm$^{-1}$, though the value, and sign, is highly sensitive to details of the electronic structure \cite{SI}. The origin of this large AHC is due to the dominant flat 5$f$ uranium bands creating a remarkable number of topological features close to $E_F$. Fig. \ref{figbasic}C shows the location of a subset of the Weyl nodes in the Brillouin zone that are within $\pm$ 60 meV of the Fermi level. There are at least 148 Weyl points, which is by far larger than other Weyl materials. There exist both type I and type II Weyl nodes, as well as a few that are proximate to a Lifshitz transition between the two types (See Fig. \ref{figbasic}D). Such a "quantum-critical" Weyl node was argued to be the origin of the large Nernst response in Co$_2$MnGa \cite{sakai2018giant}. We additionally find two topological nodal lines close to $E_F$ within the $\sigma_z$ mirror plane (Fig. 1C). Such nodal lines arise at the crossing of bands belonging to distinct irreducible representations of the mirror plane point group and are protected against perturbations by the existence of mirror symmetry. While these nodal lines are significant sources of Berry curvature, they will not contribute to $\Omega_{xy}^z$. We note that first principles calculations of the electronic structure of uranium compounds is notoriously challenging, but the main observation of an incredibly large number of Weyl nodes and Berry curvature found here is independent of the details of the electronic structure (see \cite{SI}).

%%%%%%%%%Figure2$$$$$$$$$$$
\begin{figure}[h]
	\includegraphics[width=\linewidth]{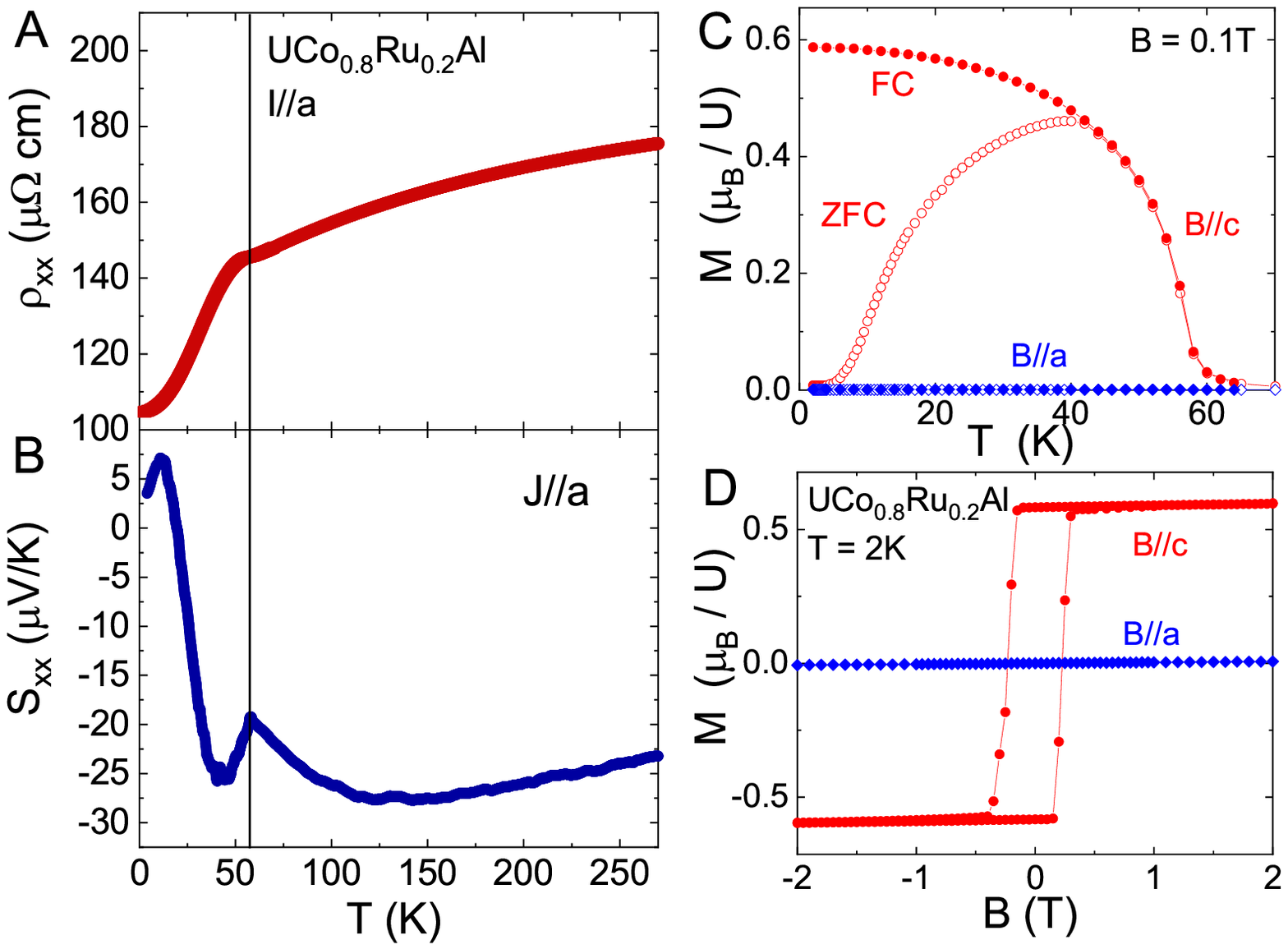}
	\caption{\label{figRSM} 
		{ \textbf{Longitudinal transport and magnetization properties of single crystalline UCo$_{0.8}$Ru$_{0.2}$Al.} (A) Electrical resistivity $\rho$ vs. temperature T with the current $I \parallel a$. The vertical solid line indicates the ferromagnetic critical temperature.  (B) Seebeck coefficient $S_{xx}$ vs. T with the heat current $J \parallel a$. (C) Temperature dependence of the magnetization $M$ per formula unit for $B \parallel c$ (red) and $B \parallel a$ (blue) in an applied field $B =$ 0.1 T. (D) Field dependence of the magnetization $M(H)$ at $T$ = 2 K.  }
	}

\end{figure}
%%%%%%%%%Figure2$$$$$$$$$$$

The temperature dependence of the longitudinal resistivity $\rho_{xx}$ and the Seebeck coefficient $S_{xx}$ of a single crystal of UCo$_{0.8}$Ru$_{0.2}$Al is shown in Fig. \ref{figRSM}A and B. A kink is observed in both $\rho_{xx}$ and $S_{xx}$ at the ferromagnetic $T_C$ = 56 K. The sample is a bad metal as reflected by the large residual resistivity (105 $\mu\Omega$cm). A longitudinal conductivity $\sigma_{xx}$ of $\sim$ 9,500 $\Omega^{-1} $cm$^{-1}$, which falls within the moderately dirty regime where $\sigma_{xx}$ = 3 $\times$ 10$^3$ - 5 $\times$ 10$^5$ $\Omega^{-1} $cm$^{-1}$, is the first indication that the transverse transport will be dominated by the intrinsic contribution \cite{onoda2008quantum}.  In the $T$=0 limit the $S_{xx}/T$ gives a value of 0.9 $\mu$V/K$^2$. This suggests a Fermi temperature of $\sim$470 K \cite{behnia2004thermoelectricity}, consistent with a Fermi temperature of 930 K estimated by specific heat \cite{SI}. We also note that the ratio of the experimental Sommerfeld coefficient to the computed value based on density functional theory implies a mass renormalization of $\sim$ 3 \cite{SI}.

The temperature and field dependent magnetization of UCo$_{0.8}$Ru$_{0.2}$Al is shown in Fig. \ref{figRSM}C and D. The system shows very strong Ising-type magnetic anisotropy, with the easy axis along the $c$-axis. The saturation magnetic moment of 0.59 $\mu_B$/U is consistent with a previous study \cite{andreev1997ferromagnetism} and implies that the uranium 5$f$ electrons are rather strongly hybridized with ligand electrons. A small field of $B =$ 0.1 T is enough to polarize the magnetic moments during field cooling (FC). Thus, we take the transverse responses (Hall, Nernst and Righi–Leduc effects) with FC and $B =$ 0.1 T conditions, which is small enough to render the ordinary contributions as negligible.

%%%%%%%%%Figure3$$$$$$$$$$$
 \begin{figure}[h]
 	\includegraphics[width=\linewidth]{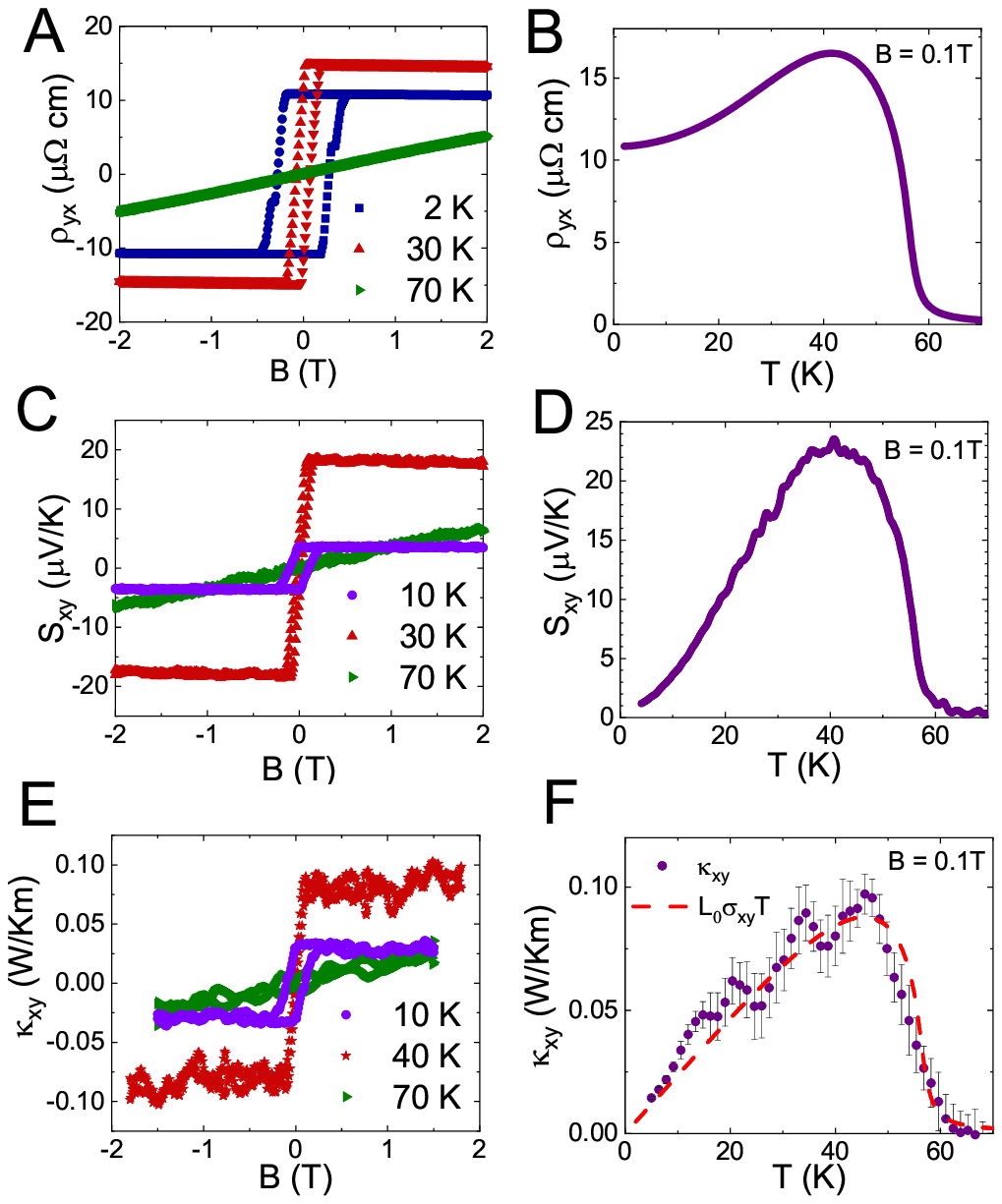}
 	\caption{\label{figtrans} 
 		{ \textbf{Colossal anomalous transverse responses from UCo$_{0.8}$Ru$_{0.2}$Al with $B \parallel c$.} (A)(B) Hall effect $\rho_{xy}$ as a function of magnetic field $B$ and temperature $T$, respectively.  (C)(D) Nernst effect $S_{xy}$ as a function of $B$ and $T$, respectively. (E)(F) Thermal Hall (Righi–Leduc) effect $\kappa_{xy}$ as a function of $B$ and $T$, respectively. The red dashed line in (F) is the expected thermal Hall conductivity based on the electrical Hall conductivity if the Wiedemann-Franz law is satisfied. All temperature dependent data were taken with $B$ = 0.1 T.}
 	}
 \end{figure}
 %%%%%%%%%Figure3$$$$$$$$$$$

The field and temperature dependence of the Hall, Nernst and Righi–Leduc effects are shown in Figs. \ref{figtrans} A-F. For the field dependence, they all show a step-like behavior similar to the magnetization. The saturated response at high fields in the ordered state confirms that the transverse effects are dominated by the anomalous contribution and the ordinary component is negligibly small. Consequently, the transverse response discussed in the remainder of the paper implicitly refers to the anomalous component. The vanishing transverse response as the temperature exceeds $T_C$ is also consistent with this picture.

The anomalous Hall effect of UCo$_{0.8}$Ru$_{0.2}$Al is found to be very large, as shown in Figs. \ref{figtrans}A and \ref{figtrans}B.  At $T=40$ K $\rho_{xy}$ is -16 $\mu\Omega$cm, which decreases to -11 $\mu\Omega$cm at $T=2$ K, corresponding to a conductivity of $\sigma_{xy} = $ 980/$\Omega$cm. These values are similar to that of Co$_2$MnGa and Co$_3$Sn$_2$S$_2$, and among the largest values for ferromagnetic metals \cite{sakai2018giant,liu2018giant}.

Next, we display our primary result - a colossal anomalous Nernst effect. While the anomalous Hall resistivity is as large as in Co$_2$MnGa and Co$_3$Sn$_2$S$_2$, more surprisingly, the anomalous Nernst effect is found to be colossal in UCo$_{0.8}$Ru$_{0.2}$Al, reaching 23 $\mu$V/K at $T =$ 40 K (Fig. \ref{figtrans} C and D). To the best of our knowledge, this is approximately 3-4 times larger than the current record value of 6-8 $\mu$V/K held by Co$_2$MnGa \cite{sakai2018giant}. Furthermore, in UCo$_{0.8}$Ru$_{0.2}$Al the magnitude of $S_{xy}$ at 40 K is comparable to the Seebeck coefficient. Consequently, the anomalous Nernst angle $\tan(\theta_N) = |S_{xy}/S_{xx}|$ reaches 0.92 at 41 K, where both $S_{xx}$ and $S_{xy}$ have extrema with  values of -25.5 $\mu$V/K and 23.5  $\mu$V/K, respectively. 

Fig. \ref{figtrans} E and F displays the anomalous thermal Hall conductivity $\kappa_{xy}$ of UCo$_{0.8}$Ru$_{0.2}$Al as a function of field and temperature. To check the validity of the anomalous Wiedemann-Franz (WF) law $L_0=\frac{\kappa_{xy}}{\sigma_{xy}T}$ where $L_0=\frac{\pi^2k_B^2}{3e^2}$ = 2.44 $\times$ 10$^{-8}$ W$\Omega$/K$^2$, we compared $\kappa_{xy}$ with $L_0\sigma_{xy}T$ (Fig. \ref{figtrans} F). We find that they overlap each other very well and thus the WF law is valid. The validity of a transverse WF law indicates that the anomalous transverse entropy and charge flow are governed by the intrinsic contribution from the Berry curvature \cite{li2017anomalous,xu2020finite}.

{\bf Discussion}

\begin{figure}[h]
 	\includegraphics[width=\linewidth]{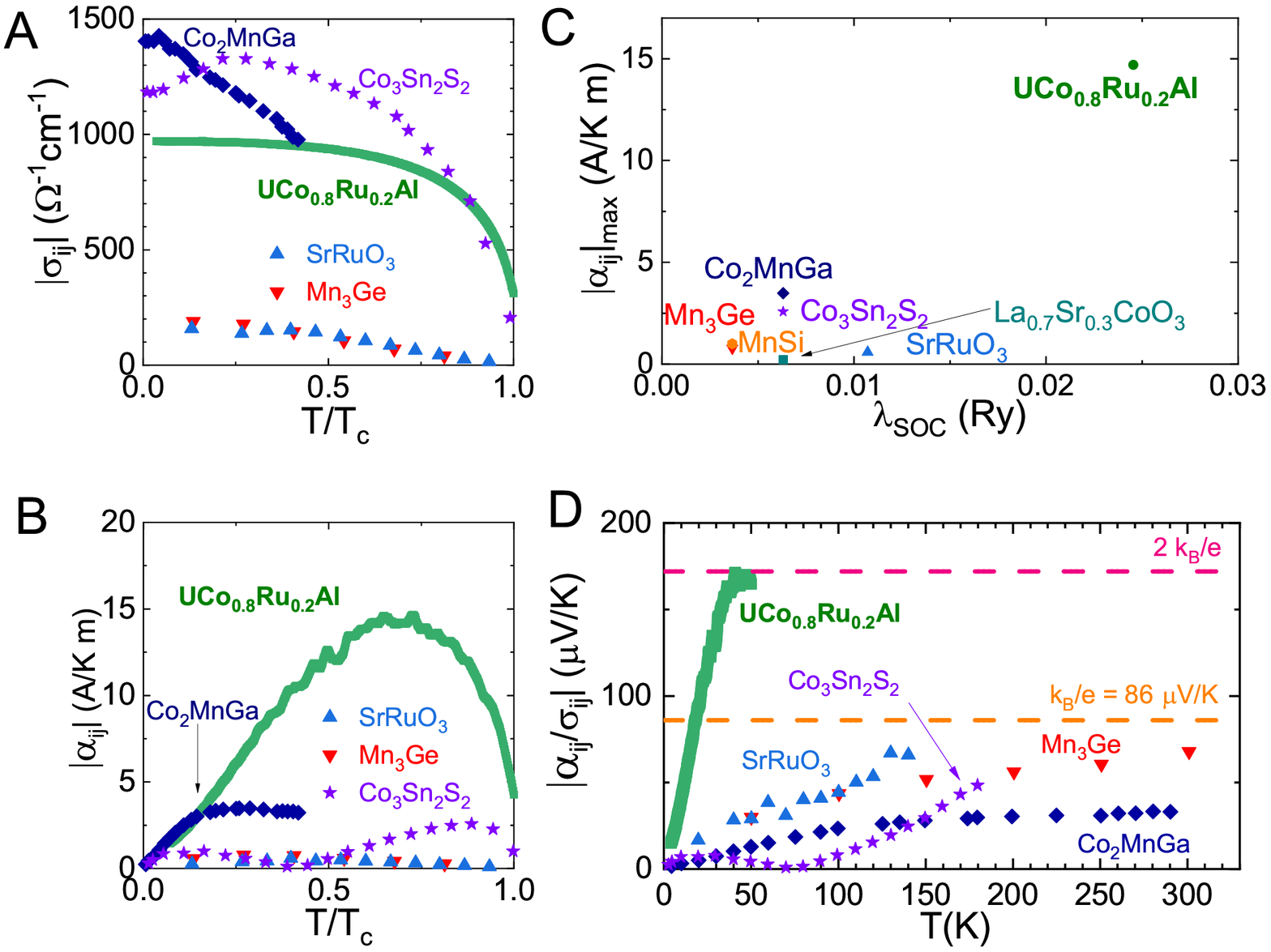}
 	\caption{\label{figAHCAPC} 
 		{\textbf{Anomalous Hall and Peltier conductivity of various magnets.} (A) Anomalous Hall conductivity $\sigma_{ij}$ and (B) Anomalous Peltier conductivity $\alpha_{ij}$ as a function of normalized temperature $T/T_c$ (green line). For comparison, other materials with large anomalous contributions are also shown. (C) The maximum value of $\alpha_{ij}$ for various compounds compared with the strength of the spin-orbit coupling of the magnetic element \cite{shanavas2014theoretical}. The compounds include UCo$_{0.8}$Ru$_{0.2}$Al (this work, green circle), Co$_2$MnGa (blue diamond) \cite{xu2020anomalous}, Co$_3$Sn$_2$S$_2$ (purple star) \cite{ding2019intrinsic}, SrRuO$_3$\cite{miyasato2007crossover} (cyan upper triangle), LSCO (light green square)  \cite{miyasato2007crossover}, Mn$_3$Ge (red lower triangle) \cite{xu2020finite},and  MnSi (orange hexagon) \cite{hirokane2016longitudinal}. (D) Plot of $\alpha_{ij}$/$\sigma_{ij}$ versus $T$. Orange and pink dashed lines indicate $k_B/e$ and 2 $k_B/e$, respectively.}
 	}
 \end{figure}

The anomalous Hall conductivity as a function of $T/T_C$ for a variety of materials with large transverse responses in heat and charge \cite{xu2020anomalous,miyasato2007crossover,ding2019intrinsic,xu2020finite} is shown in Fig. \ref{figAHCAPC} A. The AHC of UCo$_{0.8}$Ru$_{0.2}$Al is almost temperature-independent below $T$ = 0.5 $T_C$. This implies that the AHC is independent of scattering rate, which is additional support that the intrinsic contribution dominates the transverse response in this system. At $T$ = 2 K, $|\sigma_{xy}|$ reaches 980/$\Omega$cm in UCo$_{0.8}$Ru$_{0.2}$Al, which is comparable to Co$_2$MnGa, one of the largest intrinsic AHC known. The AHC from SrRuO$_3$ and Mn$_3$Ge are almost one order of magnitude smaller than that of UCo$_{0.8}$Ru$_{0.2}$Al. The anomalous Peltier conductivity $\alpha_{xy}$ as a function of $T/T_C$ for these materials is shown in Fig. \ref{figAHCAPC} B. The Peltier coefficient is the sum of two contributions $\sigma_{xx}S_{xy}$ and $\sigma_{xy}S_{xx}$. In UCo$_{0.8}$Ru$_{0.2}$Al, $\sigma_{xx}S_{xy}\gg \sigma_{xy}S_{xx}$ (see SI). In sharp contrast to the fact that $\sigma_{xy}$ of UCo$_{0.8}$Ru$_{0.2}$Al is comparable to that of Co$_2$MnGa and Co$_3$Sn$_2$S$_2$, $\alpha_{xy}$ of UCo$_{0.8}$Ru$_{0.2}$Al is roughly 3-4 times larger than that of Co$_2$MnGa, the current record holder for $\alpha_{xy}$, and Co$_3$Sn$_2$S$_2$. $\alpha_{xy}$ of SrRuO$_3$ and Mn$_3$Ge are significantly smaller.

%%%%%%%%%Figure4$$$$$$$$$$$
 %%%%%%%%%Figure4$$$$$$$$$$$

Why is $\alpha_{ij}$ so large in UCo$_{0.8}$Ru$_{0.2}$Al? In Fig. \ref{figAHCAPC} C we plot the maximal value of $\mid\alpha_{ij}\mid$ vs $\lambda_{SOC}$ for various compounds. $\lambda_{SOC}$ is estimated from atomic calculations of the magnetic element \cite{herman1963atomic,shanavas2014theoretical}. Despite this very rough estimate for the strength of the SOC in these materials, Fig. \ref{figAHCAPC} C supports our original conjecture that a large spin-orbit coupling increases the possibility for more band inversions, which can create a large Berry curvature contribution to anomalous transverse response functions \cite{xiao2010berry}. 

A large SOC alone, however, is insufficient to explain the remarkable properties of UCo$_{0.8}$Ru$_{0.2}$Al, because the maximal contribution from any pair of bands to the AHE is $e^2/hc$,  where $c$ is the $c$-axis lattice constant, and in this limit one expects a vanishing ANE. Instead our electronic structure calculations shown in Fig. \ref{figbasic} reveal that there are many bands in the vicinity of the Fermi energy. This is a consequence of the nature of the $5f$ electrons in uranium. Furthermore, electronic correlations will further renormalize the bands leading to smaller effective Fermi energies, which is known to enhance thermoelectric properties \cite{behnia2004thermoelectricity}. Notably, the Sommerfeld coefficient of the heat capacity $\gamma$ is significantly larger for UCo$_{0.8}$Ru$_{0.2}$Al ( 41 mJ/mol K$^2$) compared to  Co$_2$MnGa (12 mJ/mol K$^2$) \cite{sakai2018giant}. Thus, it is the combination of the small effective Fermi energy and the large SOC that leads to multiple topological features within the energy window set by the SOC and that enables the remarkably large values of both $\sigma_{ij}$ and $\alpha_{ij}$ in UCo$_{0.8}$Ru$_{0.2}$Al.

It is interesting to ask, which specific features of the electronic structure are most responsible for the large anomalous Nernst effect in UCo$_{0.8}$Ru$_{0.2}$Al. UCo$_{0.8}$Ru$_{0.2}$Al crystallizes in the non-symmorphic space group 189. From basic symmetry considerations this space group has been argued to possess a variety of interesting topological features, including Weyl nodes\cite{huang2015weyl,soluyanov2015type,wan2011topological,weng2015weyl}, nodal lines \cite{burkov2011topological,du2017cate,kim2015dirac,yu2015topological}, and triple points\cite{zhu2016triple,ivanov2019monopole}. All of these features appear in our electronic structure calculations \cite{SI}. Notably, one set of critically tilted Weyl points lies 25 meV above $E_f$, corresponding to a doping of $x\sim 0.07$ in the rigid band picture. For this set of Weyl nodes we compute values of $\sigma_{xy} \sim 1200 $ ($\Omega$ cm)$^{-1}$ and $\alpha_{xy} \sim 20 $ A(K m)$^{-1}$ at that energy, which may be shifted as a consequence of electronic correlations.  Another possibility is that electronic correlations pin the Weyl nodes to the chemical potential as has been suggested in some Weyl-Kondo semimetals \cite{lai2018weyl}. Additional measurements, such as ARPES, will be required to confirm the identity of the most relevant topological features in UCo$_{0.8}$Ru$_{0.2}$Al. 

Finally, we consider the relationship between $\sigma_{ij}$ and $\alpha_{ij}$. From the Mott relation, which is valid in the $T$ = 0 limit, a large $\alpha_{ij}/T$ implies that $\sigma_{ij}$ can be enhanced by appropriately tuning the chemical potential. As temperature is increased the relationship between $\sigma_{ij}$ and $\alpha_{ij}$ is less straight forward. Behnia \textit{et al.} have argued, however, that the ratio of $\alpha_{ij}$/$\sigma_{ij}$ is bounded by the ratio of natural units $k_B$/$e$ = 86 $\mu$V/K, which reflects the fact that the Peltier coefficient represents the transport of entropy, while the Hall coefficient represents the transport of charge \cite{xu2020anomalous}.  For UCo$_{0.8}$Ru$_{0.2}$Al $\alpha_{ij}/\sigma_{ij}$ reaches 170 $\mu$V/K at $T$ = 47 K, as shown in Fig. \ref{figAHCAPC} D. This supports the notion that multiple bands are simultaneously contributing to the large Nernst response in UCo$_{0.8}$Ru$_{0.2}$Al.

In conclusion, we have found a colossal anomalous Nernst effect in UCo$_{0.8}$Ru$_{0.2}$Al. The simultaneous presence of strong spin-orbit coupling and electronic correlations drives narrow bands leading to a gigantic Nernst response and presents an exciting blueprint for how to develop magnetic topological metals for thermoelectric applications.

{\bf Materials and Methods}

A single crystal of nominal composition UCo$_{0.8}$Ru$_{0.2}$Al was grown by the Czochralski method in a tri-arc furnace. The crystal structure was confirmed to have the P$\bar{6}2m$ (\# 189) space group with the ZrNiAl structure type. Powder and single crystal diffraction was performed to confirm the crystal structure and the phase purity of the samples. The sample was oriented and the quality was further checked by the X-ray Laue method with a photo-sensitive detector in backscattering geometry. The measured magnetic properties are in good agreement with previous studies, further confirming the sample quality.

%{\cblue Seva, we will also need details of the electronic structure calculations. The following is from a writeup from 10-14-19. Need calculation parameters such as lattice constant used.}
%To better understand the origin of the anomalous Hall and Peltier effects in the UCo$_{0.8}$Ru$_{0.2}$Al system, we computed the electronic structure using the full-potential linear muffin tin orbital (FP-LMTO) method [18]. We used a self consistent local spin density approximation (LSDA), which had previously been employed to study the complex magnetism [7-9] of the UTAl series. 

Electronic structure calculations were performed using the local spin density approximation (LSDA), yielding a ferromagnetic ground state, within the framework of the full potential linear muffin tin orbital method with spin-orbit coupling \cite{savrasov1996linear}. This approximation has been previously used to study UCoAl \cite{antonov2003electronic,kuvcera2002x} and other uranium compounds in the same structural family \cite{gasche1992theory}. Experimental lattice constants were used \cite{lam1967equiatomic}, and doping dependence was handled using a rigid band approximation.

Weyl points and other topological features were located and their topology verified in a single shot method \cite{ivanov2019monopole}. An initial coarse $\bm{k}$-grid of $30 \times 30 \times 30$ was used to locate candidate sources/sinks of Berry curvature flux, whose positions were iteratively refined by repeating the procedure on $2 \times 2 \times 2$ grids within each $\bm{k}$-cube. Additional details can be found in \cite{SI}.

Transport, thermoelectric, and thermal transport measurements were performed in the same setup. A temperature gradient was generated by a 10 k$\Omega$ chip resistor attached to one end of the sample and measured by home-made thermocouples with the calibrated table made of Fe-doped gold and chromel. The other end of the sample was attached to a sapphire substrate that was physically clamped to an oxygen-free copper coldfinger.  Conventional steady-state measurements were performed to measure the thermoelectic power and thermal conductivity. The heating power was carefully monitored so that the thermal and thermoelectric responses were linear with the heating power.

{\bf Acknowledgement:} 
We thank Priscila Rosa and Xiangang Wan for useful discussions. The experimental work was performed at Los Alamos National Laboratory under the auspices of the U.S. Department of Energy, Office of Science, Basic Energy Sciences, Materials Sciences and Engineering Division. Theoretical calculations performed by V.I. and S.Y.S. are supported by NSF DMR Grant No. 1832728. T.A. acknowledges support from a LANL LDRD Directors Postdoctoral Fellowship. {\bf Author contributions:} T.A. and F.R. designed the experiment and analyzed the data. T.A., S.M.T., and F.R. built the transport measurement setup. T.A. performed the transport and heat capacity measurements. J.D.T. carried out the magnetic measurements. V.I. and S.Y.S. performed the theoretical calculations. E.D.B. prepared the sample. All authors contributed to writing the manuscript. {\bf Competing interests:} The authors declare that they have no competing interests. {\bf Data and materials availability:} All data needed to evaluate the conclusions in the paper are present in the paper and/or the Supplementary Materials. Additional data related to this paper may be requested from the authors.

\end{document}

%\preprint{}

\title{Colossal anomalous Nernst effect in correlated noncentrosymmetric kagome ferromagnet}
\author{T. Asaba$^{1*}$, V. Ivanov$^{2}$, S. M. Thomas$^{1}$, S.Y. Savrasov$^{2}$, J. D. Thompson$^{1}$, E. D. Bauer$^1$, F. Ronning$^{3*}$}

\affiliation{
$^1$Materials Physics and Applications Division, Los Alamos National Laboratory, NM, 87545 USA \\
$^{2}$Department of Physics, University of California, Davis, CA 95616, USA \\
$^{3}$Institute for Materials Science, Los Alamos National Laboratory, NM, 87545 USA \\}
\date{\today}
%\begin{abstract}

%\end{abstract}

\maketitle
\section{Computational details}

\indent To obtain the electronic structure of doped UCo$_{0.8}$Ru$_{0.2}$Al, we performed a series of electronic structure calculations within the framework of the full potential linear muffin tin orbital method with spin-orbit coupling \cite{savrasov1996linear}, using the local spin density approximation (LSDA), which has been previously employed to study uranium compounds with the ZrNiAl-type structure \cite{gasche1992theory,gasche1995ground,gasche1995ground2}.\\
\indent To assess the effect of chemical substitution on the electronic structure, we perform calculations for the parent compound UCoAl, as well as doped at the $x = 1/3, 2/3$ and $1$ levels, using the experimental lattice parameters. The UTAl (T=Co,Ru) structure has two inequivalent sites for atom T, T$_1$ and T$_2$, with Ru preferentially occupying the two T$_2$ sites \cite{chang2000magnetism,chang2000crystallographic}. Therefore calculations on the chemically ordered compounds UCo$_{2/3}$Ru$_{1/3}$Al, UCo$_{1/3}$Ru$_{2/3}$Al with Ru at the T$_2$ site will give a representative picture of the doping effects on the electronic structure of UCo$_{1-x}$Ru$_x$Al. The resulting band structures are shown below.\\

\begin{figure}[hb]
	\centering
	\begin{subfigure}[b]{0.475\textwidth}
		\centering
		\caption{UCoAl}
		\includegraphics[width=\textwidth]{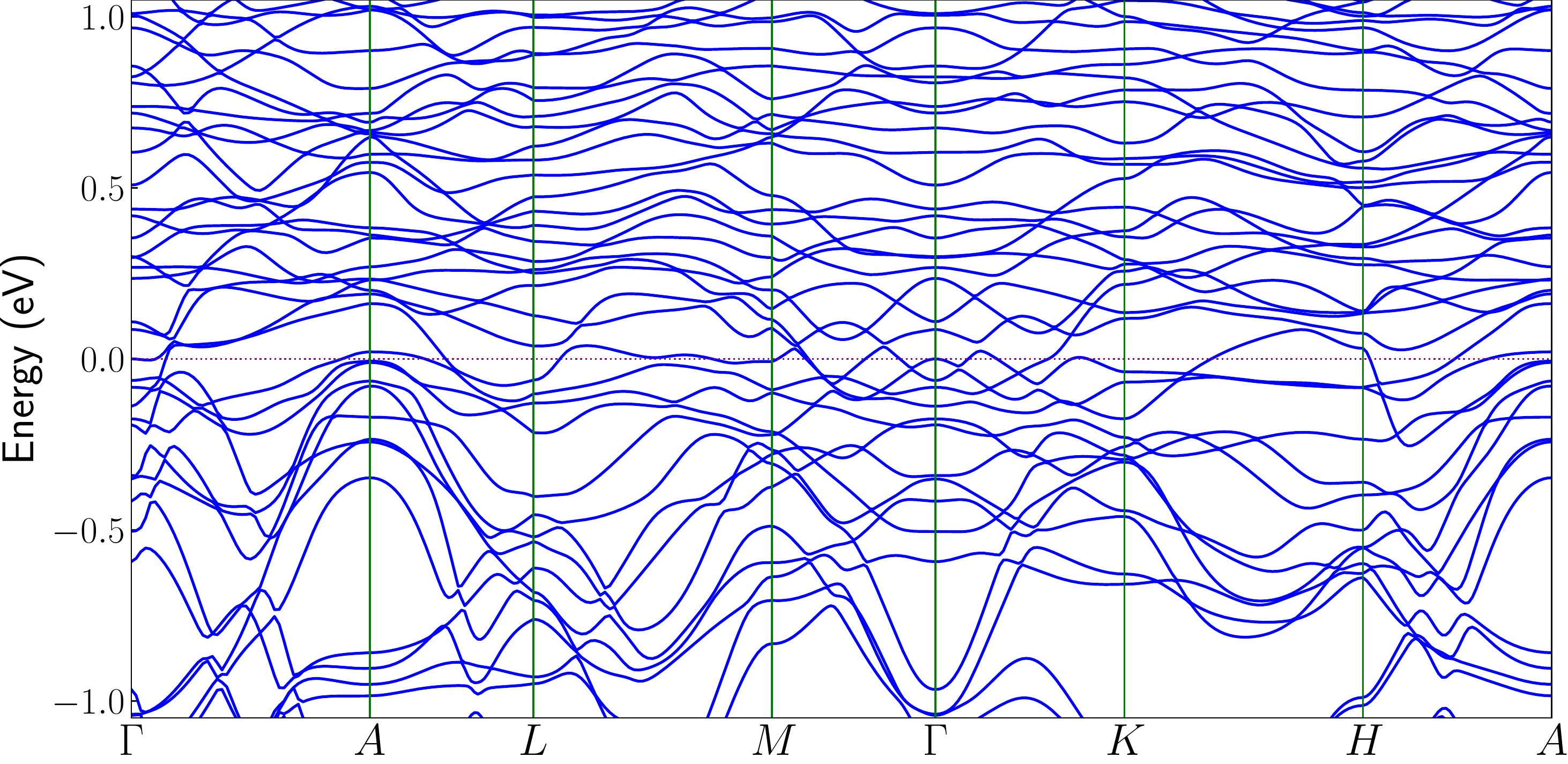}
	\end{subfigure}
	\hfill
	\begin{subfigure}[b]{0.475\textwidth}  
		\centering
		\caption{UCo$_{2/3}$Ru$_{1/3}$Al}
		\includegraphics[width=\textwidth]{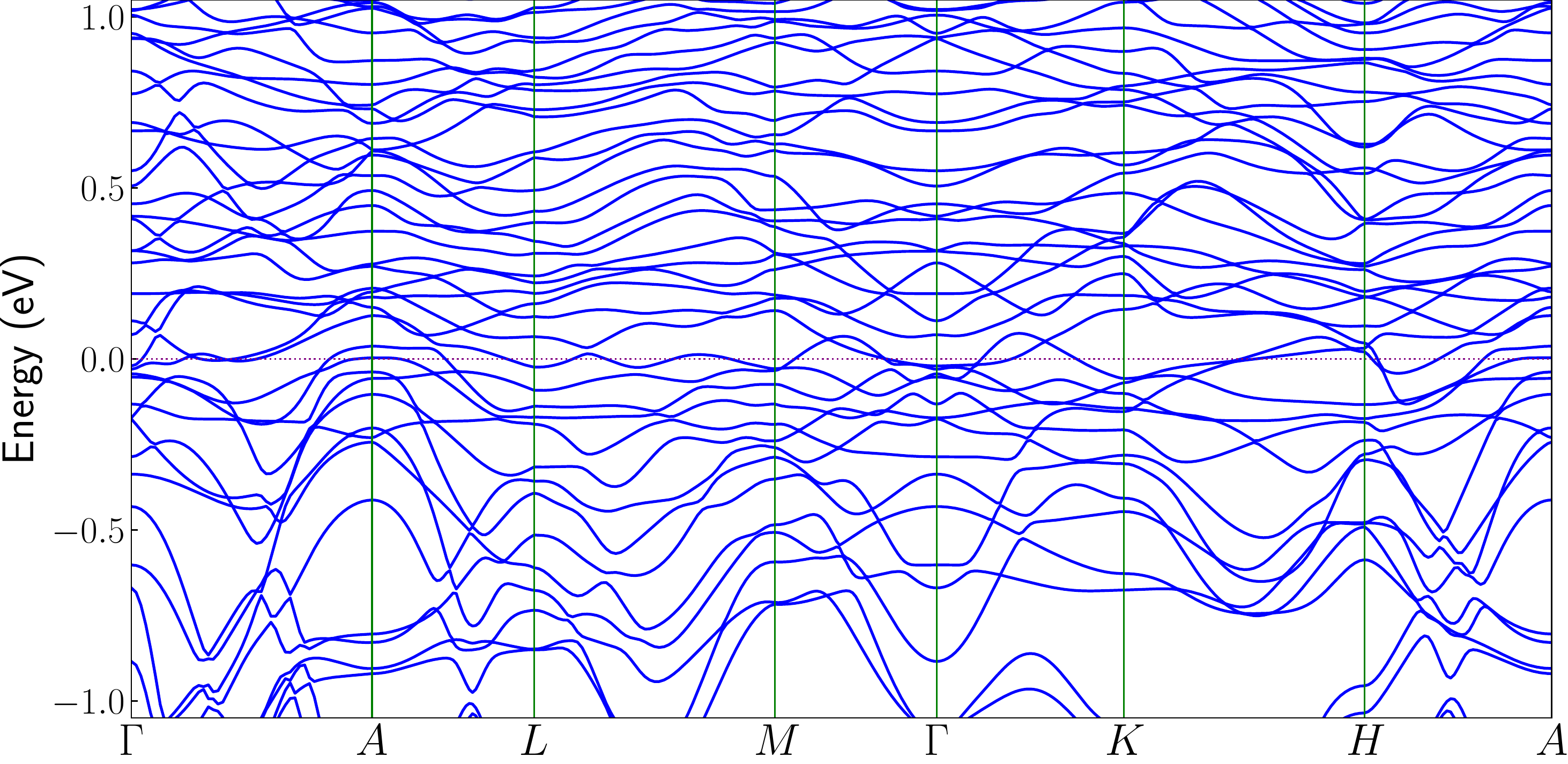} 
	\end{subfigure}
	\vskip\baselineskip
	\begin{subfigure}[b]{0.475\textwidth}   
		\centering
		\caption{UCo$_{1/3}$Ru$_{2/3}$Al}
		\includegraphics[width=\textwidth]{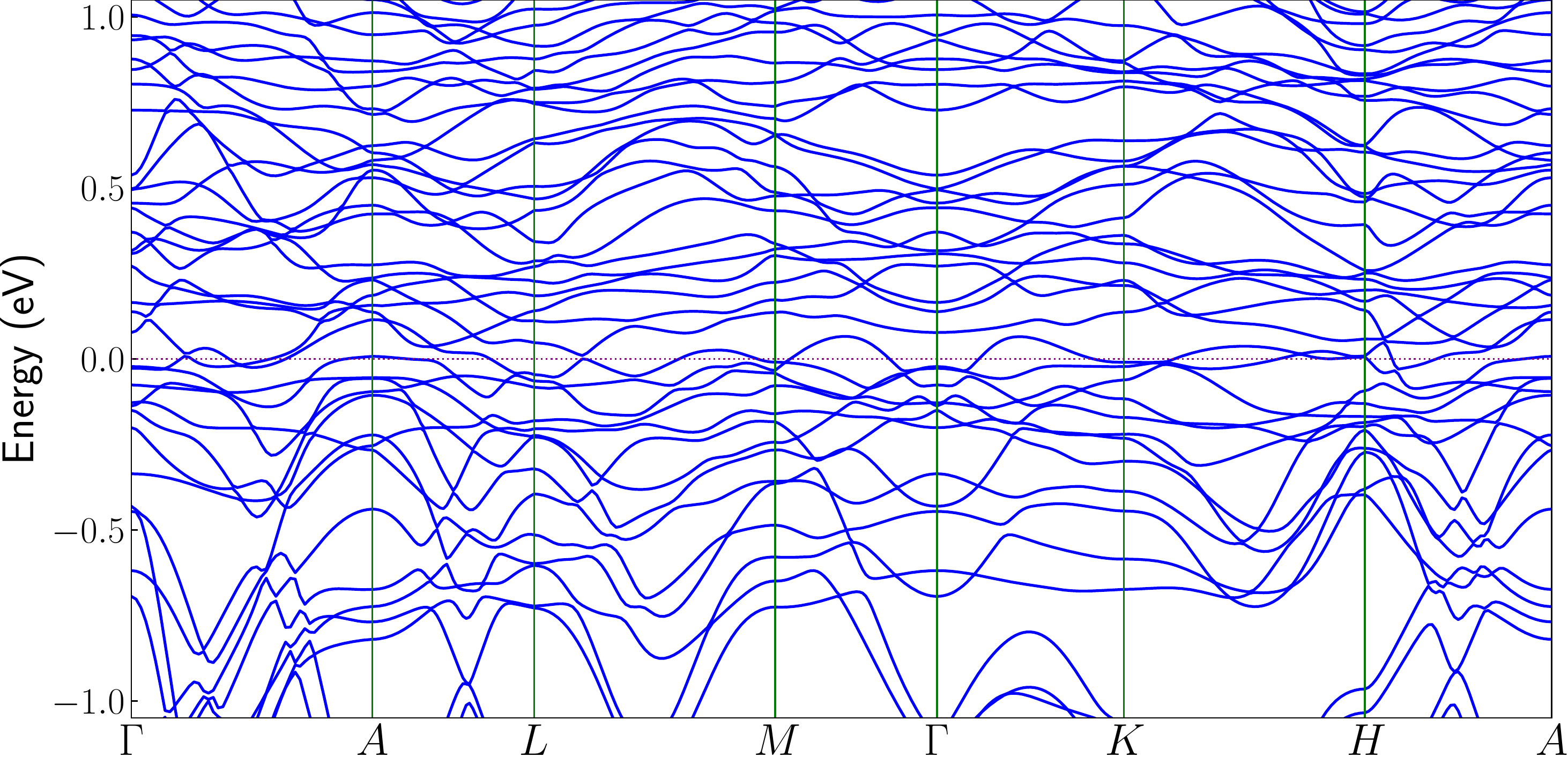}
	\end{subfigure}
	\hfill
	\begin{subfigure}[b]{0.475\textwidth}   
		\centering
		\caption{URuAl}
		\includegraphics[width=\textwidth]{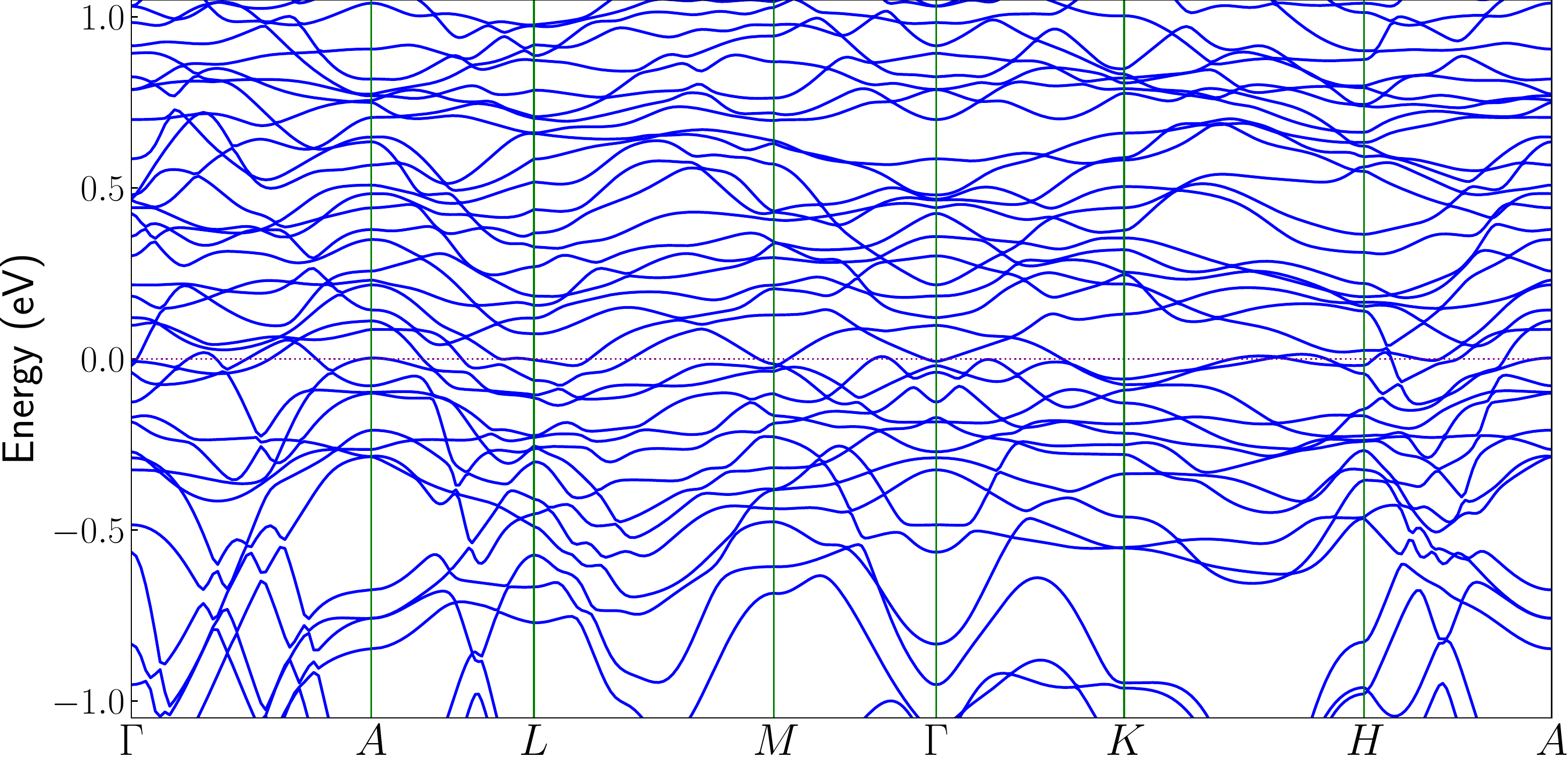}
	\end{subfigure}
	\caption{\textbf{Calculated band structures of chemically ordered members of the UCo$_{1-x}$Ru$_x$Al series.} Computed band structures for the stoichiometric compounds in the UCo$_{1-x}$Ru$_x$Al series, with $x = 0, 1/3, 2/3, 1$. $E_F=0$ is the Fermi level corresponding to the $x=0.2$ doping.} 
	\label{ucra-bands}
\end{figure}

\indent Figure \ref{ucra-bands} shows that as Ru is substituted for Co, the band structure remains largely unchanged, aside from a downward shift in the Fermi level due to the removal of electrons. These calculations indicate that a rigid band approximation for doped UCo$_{1-x}$Ru$_x$Al is appropriate. Consequently, to approximate the electronic structure of UCo$_{0.8}$Ru$_{0.2}$Al, we use the electronic structure of pure UCoAl with a rigid band shift up of 41.04 meV corresponding to the $x=0.2$ doping level. This will be used consistently throughout the manuscript and the supplement. \\
\indent The uranium $f$-orbitals reside on a kagome lattice. Thus, the effects of frustrated hopping on such a lattice and the small spatial extent of the 5f orbitals will be intermixed. Identifying specific bands associated with the kagome structure is further complicated by the multiorbital nature of uranium, the presence of ligand orbitals, and the three dimensionality of the structure. Nevertheless, multiple flat bands can be observed in the bandstructure plots of Fig. \ref{ucra-bands}. One would expect the flat bands near the Fermi energy to be renormalized by electronic correlations present in this system. \\
\indent For UCoAl, the computed orbital $\mu_l = 1.19 \mu_B$ and spin $\mu_s = -0.98 \mu_B$ magnetic moments, which are similar to those found in prior calculations \cite{kuvcera2002x}. Experimentally UCo$_{1-x}$Ru$_x$Al is paramagnetic for $x$ = 0, developing a static magnetic moment in the doping range $x = 0.005 - 0.78$, with a maximum uranium magnetic moment $\mu_U = 0.6 \mu_B$ around $x = 0.3$\cite{andreev1996onset}. Therefore, with the appropriate adjustment of Fermi level, our calculated electronic structure for UCoAl is a reasonable model to compare with experimental observations for doped UCo$_{0.8}$Ru$_{0.2}$Al. It is important to note that our model is only valid at energies corresponding to dopings within the magnetic range, and not for the paramagnetic end points.  We further emphasize that while details of the actual electronic structure may vary, the main observation of an abundance of topological features close to $E_F$ will be insensitive to all detailed variations of the electronic structure.\\

\section{Identification of Topological Features} 

\begin{figure}[h]
	\centering
	\includegraphics[width=0.5\textwidth]{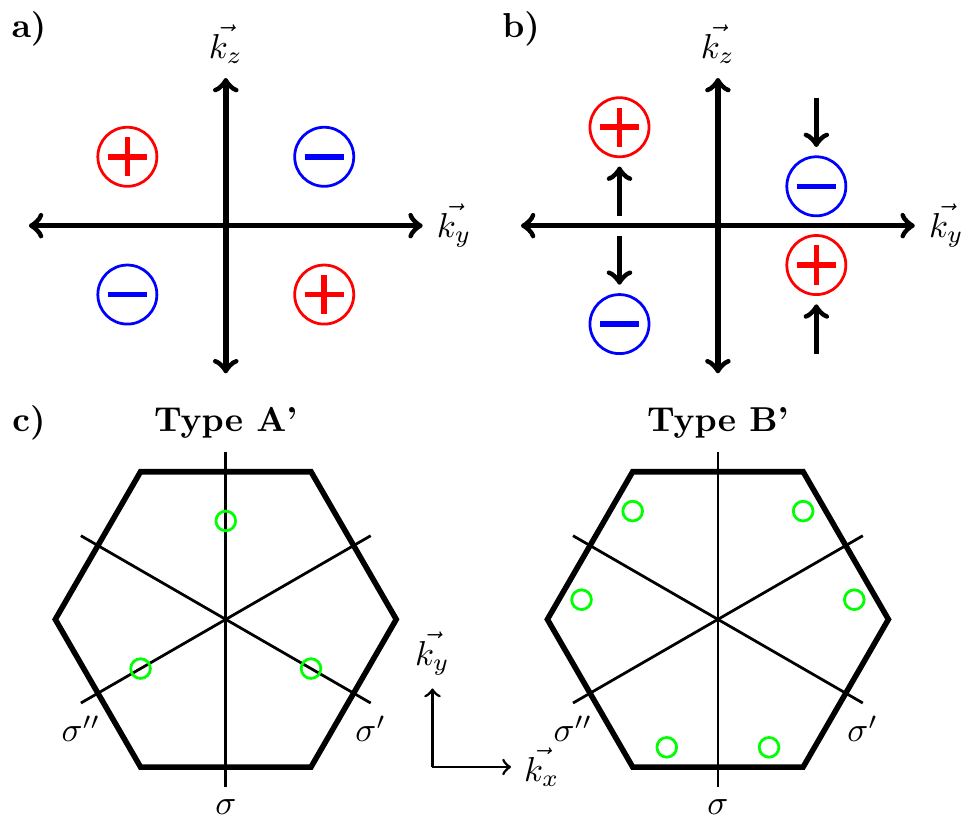}
	\caption{\textbf{Schematics showing how breaking $\vec{k} \leftrightarrow -\vec{k}$ symmetry splits Weyl points in the ZrNiAl-type structure.} a) 4 Type-A Weyl points in UCo$_{1-x}$Ru$_{x}$Al in the absence of magnetism. b) Turning on the magnetic field breaks $\textbf{k}\leftrightarrow -\textbf{k}$ symmetry, splitting the original set of Weyl points into two sets of Type A' or Type B'. c) Symmetries of Type A' and Type B' Weyl points in UCo$_{1-x}$Ru$_{x}$Al, as seen from above.}
	\label{weylsym} 
\end{figure}

\indent To understand the origin of the large anomalous Hall- and anomalous Nernst effects in UCo$_{0.8}$Ru$_{0.2}$Al, we perform a previously developed mining procedure \cite{ivanov2019monopole} to find topological features in the electronic structure. Because the topological features closest to the Fermi level will be the most relevant to the transport properties, we scanned the range of energies $E_F \pm 60 meV$ for topological features.\\
\indent We divide the Brillouin zone (BZ) into an initial coarse $k$-grid of $30 \times 30 \times 30$ divisions, and compute the Berry curvature flux through the surface of each $k$-cube. This grid is subsequently refined by iteratively repeating the search procedure on a $2 \times 2 \times 2$ grid within each $k$-cube until the desired precision is achieved. This allows us to find the locations of sources/sinks of Berry curvature flux, simultaneously confirming the topological nature of the features as well as their positions in $k$-space. Our procedure reveals a number of topological features in UCo$_{0.8}$Ru$_{0.2}$Al, including Weyl points, nodal lines, and triple points. The triple points \cite{zhu2016triple} only exist within the paramagnetic state, and therefore, will not be discussed in this manuscript. Both Weyl nodes and nodal lines exist in the ferromagnetic state. \\
%TODO cite
%
\indent The Weyl points we identify in UCo$_{0.8}$Ru$_{0.2}$Al are summarized in Table \ref{weyl-data}, and fall into one of three general classes. The main two types of Weyl points we term Weyl-A' and Weyl-B', due to their relationship to the Weyl-A and Weyl-B types of Weyl points we have identified for compounds with the ZrNiAl-type structure \cite{ivanov2019monopole}.\\
\begin{table}[hb]
	\centering
	\begin{tabular}{ c  c  c | c c c c c}
		\hline
		\hline
		Band & Location & Type & \# & $\delta k_z$ & $v$ (Ry/$\hat{k}$) & $C$ (Ry/$\hat{k}$) & E(meV)\\% & E(Ry)  \\
		\hline
		71 & ( 0.00000, 0.33057, 0.02756) & Weyl-A' & 6	& 0.0030000 &-0.0418333	& 0.0183333	&-63\\%& 0.7900918573 \\
		71 & ( 0.00000,-0.53900, 0.09977) & Weyl-A' & 6 & 0.0268000 &-0.0549254	& 0.0107090	&-30\\% & 0.7925359845 \\
		71 & ( 0.00000, 0.15877, 0.12084) & Weyl-A' & 6 & 0.0174000 &-0.0427011	&-0.0177299	&-15\\% & 0.7936480000 \\
		71 & ( 0.00000,-0.45758,-0.12498) & Weyl-A' & 6 & 0.0317400 & 0.0811122	&-0.0017171	&-22\\% & 0.7931363466 \\
		71 & ( 0.00000, 0.55407, 0.22190) & Weyl-A' & 6 & 0.0178400 &-0.0225897	& 0.0070348	& -6\\% & 0.7942962631 \\
		71 & ( 0.61459, 0.05412,-0.12708) & Weyl-B' &12 & 0.0304600 & 0.0391005	& 0.0207814	&-51\\% & 0.7917189304 \\
		71 & ( 0.41976, 0.10908,-0.42260) & Weyl-B' &12 & 0.0112000 & 0.0174554	&-0.0062946	&-32\\% & 0.7923951491\\
		71 & ( 0.00000, 0.31823,-0.45537) & Weyl-A' & 6 & 0.0112000 & 0.0132366	&-0.0018973	&-44\\% & 0.7915001452 \\
		71 & ( 0.00000,-0.38203, 0.38948) & Weyl-A' & 6 & 0.0240000 &-0.0208750	& 0.0056250	& +2\\% & 0.7948439715 \\
		%\hline
		72 & ( 0.00000, 0.38736, 0.00434) & Weyl-A' & 6 & 0.0040000 &-0.0842500	& 0.0823750	&-51\\% & 0.7909740446 \\
		72 & ( 0.00000, 0.51091,-0.01959) & Weyl-A' & 6 & 0.0078000 & 0.0377885	&-0.0377884	&+26\\% & 0.7966048845 \\
		72 & ( 0.00000,-0.07528,-0.03239) & Weyl-A' & 6 & 0.0049560 & 0.0329903	&-0.0069613	&-23\\% & 0.7930098196 \\
		72 & ( 0.27505,-0.00369, 0.03266) & Weyl-B' &12 & 0.0030000 &-0.0510833	& 0.0505833	&-32\\% & 0.7923571598 \\
		72 & ( 0.00000, 0.08053, 0.02120) & Weyl-A' & 6 & 0.0048000 &-0.0233333	& 0.0063542	&-28\\% & 0.7926306717 \\
		72 & ( 0.00000, 0.23693,-0.05635) & Weyl-A' & 6 & 0.0126000 & 0.1351587	& 0.1173413	& -1\\% & 0.7946684601  \\
		72 & ( 0.00000,-0.18377, 0.09803) & Weyl-A' & 6 & 0.0130000 &-0.1004615	& 0.0413462	&+28\\% & 0.7968054764 \\
		72 & ( 0.40132,-0.13379,-0.13770) & Weyl-B' &12 & 0.0130000 & 0.0300000	& 0.0206538	&+59\\% & 0.7990951920 \\
		72 & ( 0.00000, 0.00000, 0.22577) & $k_z$   & 2 & 0.0210000 &-0.0063095	& 0.0300238	&+57\\% & 0.7989248079 \\
		%		72 & ( 0.00000, 0.00000,-0.11234) & $k_z$   & 2 &			&			&			&-0.4\\%& 0.797707		\\
		%		72 & ( 0.43878, 0.16586, 0.38870) & Weyl-B' &12 &			&			&			&+2.9\\%& 0.79795  \\
		%		72 & ( 0.00000,-0.45434, 0.38414) & Weyl-A' & 6 &			&			&			&+14 \\%& 0.798761 \\
		%		73 & ( 0.00000, 0.00000, 0.08663) & $k_z$   & 2 &			&			&			&+13 \\%& 0.798660 \\
		\hline
	\end{tabular}
	\caption{\textbf{Weyl points of UCo$_{0.8}$Ru$_{0.2}$Al.} The first column gives the lower band number of two bands comprising the Weyl point. The second column gives the position of one positive charge given for each symmetry-related set of Weyl points. The third and fourth columns give the classification of the Weyl points as Weyl-A', Weyl-B', or $k_z$, and the number of symmetry-related Weyl points in the set. The remaining columns give the momentum cutoff $\delta k_z$, velocities $v$ and $C$, and energy $E$, of each Weyl point. Wavevectors $\bm{k}$ are given in units of $2\pi/a,2\pi/a,2\pi/c$.}
	\label{weyl-data}
\end{table}
\indent For reference, Weyl points belonging to type Weyl-A form six pairs found along the $\Gamma-M$ line and are separated along the $k_z$ direction. Weyl-B points are instead 12 pairs which are found in sets of four, symmetrically displaced from the $\Gamma-K$ line. The reason for 12 Weyl points being the minimum number can be understood in terms of a symmetry argument. While Weyl points can arise due to the absence of either time-reversal $\mathcal{T}$ or inversion $\mathcal{I}$ symmetries, here they exist due to broken inversion. Now, while $\mathcal{T}(\bm{k}) = - \bm{k}$, the topological charge of a Weyl point is invariant under $\mathcal{T}$. This means that a positively charged Weyl point located at $\bm{k}$ will have a positively charged partner at $-\bm{k}$, each of which will have a negatively charged partner across the $\sigma_z$ mirror plane as shown in Fig~\ref{weylsym}a. In combination with the three-fold symmetry of the BZ, this guarantees that each Weyl point belongs to a symmetry-related set with a number of members that is a multiple of twelve.\\
%\\	
\indent In UCo$_{1-x}$Ru$_{x}$Al, the uranium magnetic moments break $\mathcal{T}$-symmetry on top of the already absent inversion symmetry of the lattice. This means that the inversion symmetry $\textbf{k}\leftrightarrow -\textbf{k}$ in $\bm{k}$-space is fully broken, splitting the Weyl A and Weyl B types each into two sets of Weyl A' or Weyl B' types, with 6 and 12 members respectively (Fig~\ref{weylsym}c). This can be understood as a direct consequence of the Zeeman-like effect shifting the bands, causing the $k_z$-separation between Weyl point partners to increase or decrease (Fig~\ref{weylsym}b). A similar mechanism can create lone pairs of Weyl points along the $\Gamma-A$ axis, separated only along the $k_z$ direction \cite{ivanov2018correlation}. Since this sort of Weyl point is pinned to the $k_z$-axis, application of point group symmetries does not yield any new symmetry-related members. Therefore these sets have only two members, which is the minimum for materials with broken $\mathcal{T}$-symmetry.\\

\begin{figure}[H]
	\centering
	\begin{subfigure}[b]{0.4\textwidth}
		\centering
		\includegraphics[width=\textwidth]{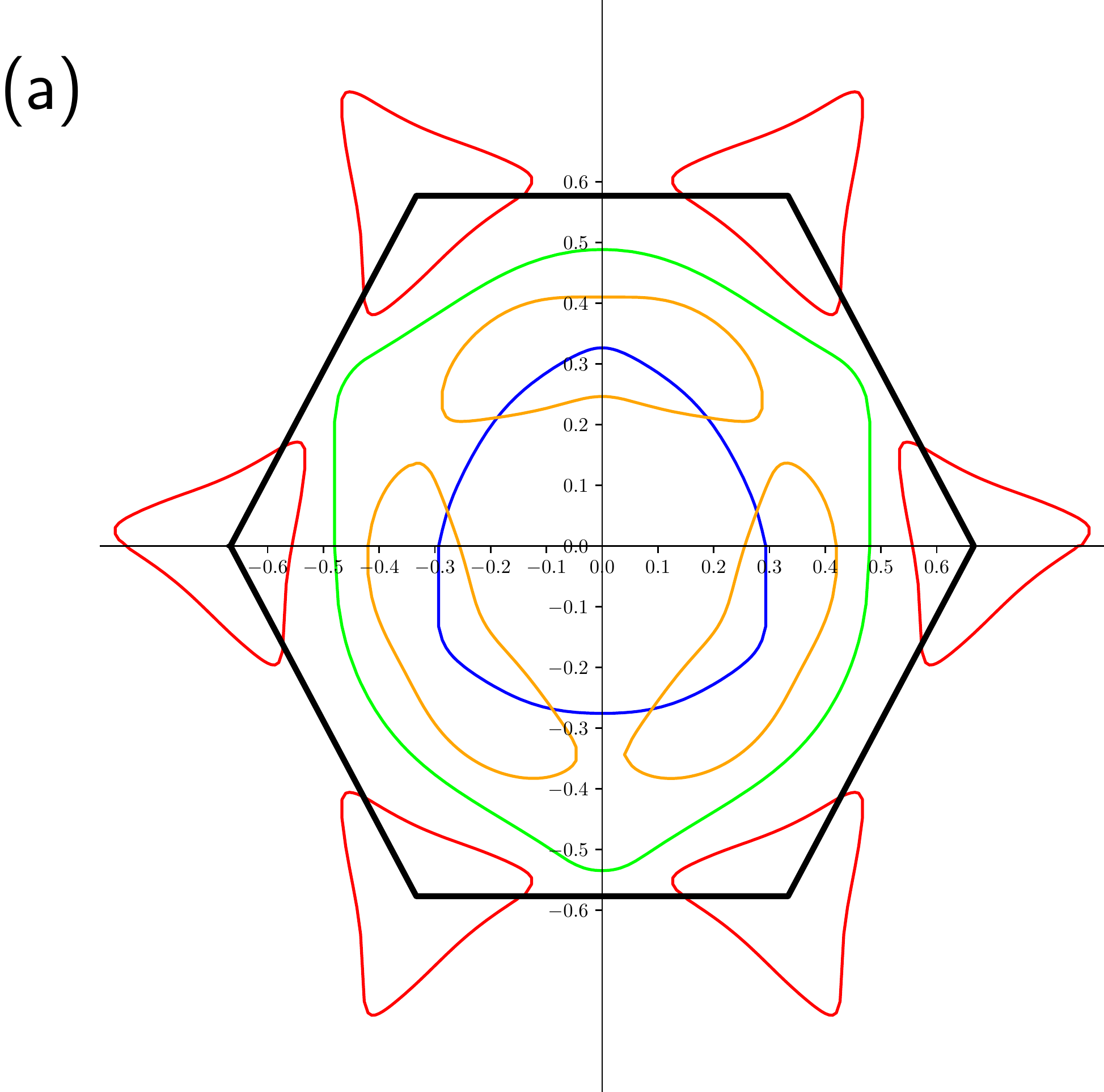}
	\end{subfigure}
	\hspace{1em}%\hfill
	\begin{subfigure}[b]{0.4\textwidth}  
		\centering
		\includegraphics[width=\textwidth]{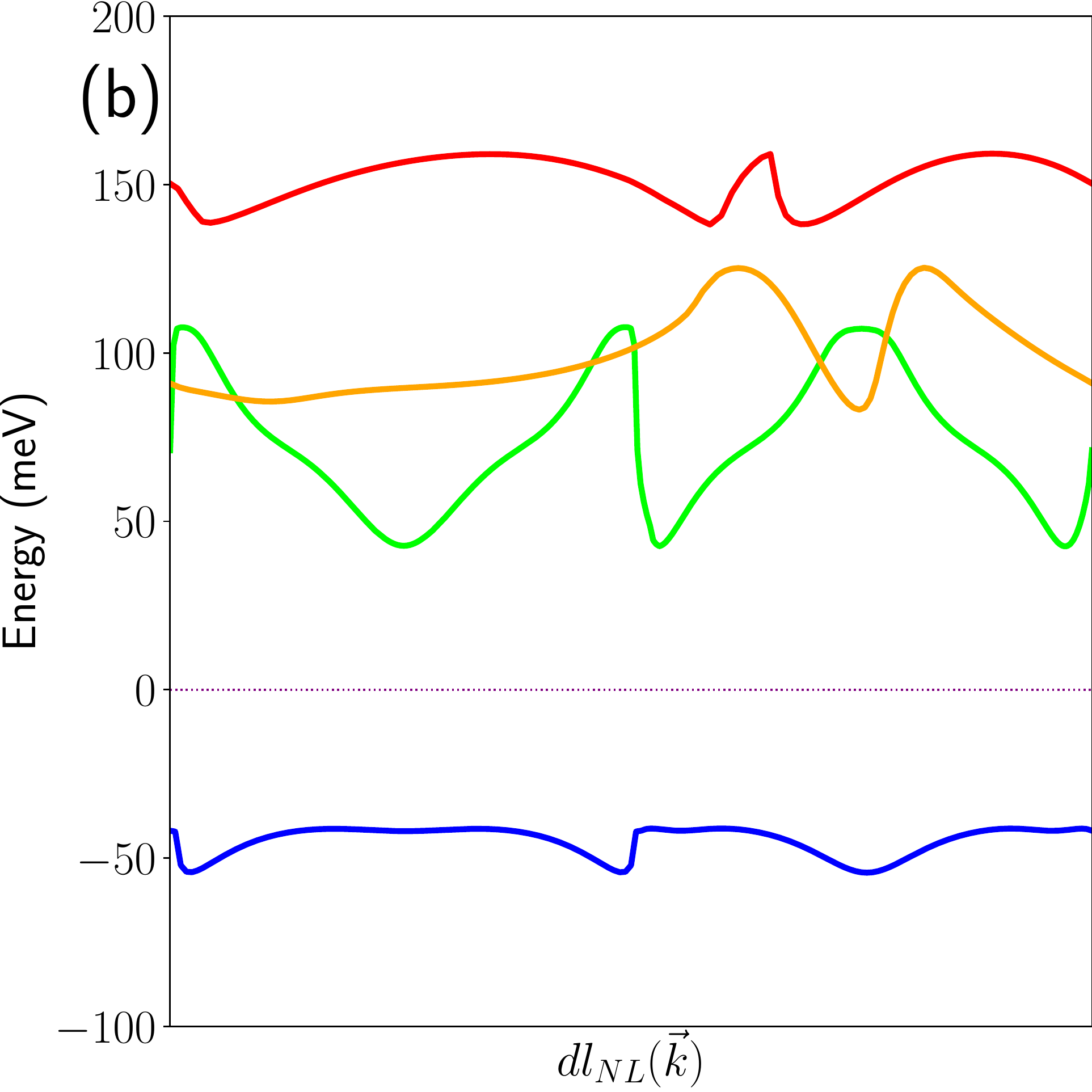} 
	\end{subfigure}
	\hfill
%	\begin{subfigure}[b]{0.3\textwidth}   
%		\centering
%		\includegraphics[width=\textwidth]{UCRA-tp.pdf}%
%	\end{subfigure}
	\caption{\textbf{Nodal lines of UCo$_{0.8}$Ru$_{0.2}$Al.} a) Nodal lines within the $\sigma_z$ plane of UCo$_{0.8}$Ru$_{0.2}$Al. b) Nodal lines plotted as a function of energy. } 
	\label{ucra-nl-tp}
\end{figure}
\indent In addition to the Weyl points we find a number of topological nodal lines in the electronic structure of UCo$_{0.8}$Ru$_{0.2}$Al, shown in Figure~\ref{ucra-nl-tp}a. This type of nodal line can arise at the intersection of two bands belonging to two different irreducible representations of the mirror plane point group. In UCo$_{1-x}$Ru$_{x}$Al, these nodal lines have different eigenvalues when acted upon by the $\sigma_z$ mirror plane. As a result the band degeneracy at their intersection point is topologically protected against small distortions. These nodal lines are located at least 50 meV above/below the Fermi energy (Fig~\ref{ucra-nl-tp}b). While they occupy a larger part of the topological phase space than the Weyl points, they likely don't make a significant contribution to the Berry curvature at the Fermi energy of the $x=0.2$ doped case. Specifically, the nodal lines lying within the $\sigma_z$ plane will not contribute to $\Omega_{xy}^z$ and hence will not impact $\sigma_{xy}$ or $\alpha_{xy}$. \\
%TODO cite
%
%\indent The UCoAl crystal structure belongs to the inversion broken \textit{p}$\bar{6}$2\textit{m} (\# 189) space group, which can also host triple-points \cite{zhu2016triple}. Our procedure applied to a non-magnetic calculation was able to locate several pairs of such triple points (Fig~\ref{ucra-nl-tp}c). Like the nodal lines, these triple points are located far below the Fermi level, at -272 meV, -158 meV, -141 meV, and -191 meV below $E_F$, respectively. Hence, like the nodal lines, they would have a negligible effect on the Berry curvature calculation for UCo$_{0.8}$Ru$_{0.2}$Al. More importantly, the existence of these triple points requires time-reversal symmetry, which is broken in UCo$_{0.8}$Ru$_{0.2}$Al by the uranium magnetic moments. \\
%

% \pagebreak

\section{Weyl Model}

\indent The large number of topological features in UCo$_{0.8}$Ru$_{0.2}$Al result in a dense population of singularities in the Berry curvature. This means that the standard approach for computing anomalous Hall and anomalous Nernst effects would require dense $\bm{k}$-grids that are too large to be computationally tractable. Instead, we take a pragmatic approach by using a solvable model for each pair of identified Weyl points, allowing us to practically compute the anomalous Hall and Nernst effects.

\indent We can construct an effective model using the positions of the Weyl points in UCo$_{1-x}$Ru$_x$Al, and the velocity parameters extracted from the local band structure near the Weyl points (parameters listed in Table \ref{weyl-data}). The simplest such model simply assigns a plateau of anomalous hall conductivity $\sigma_{xy} = e^2Q/2\pi^2\hbar$ to each Weyl point within some momentum cutoff range. However, several of the Weyl points are strongly tilted, which affects their contribution to the AHC, so we will use an extended model which takes this tilt into account \cite{zyuzin2016intrinsic}:
%Link: https://link.springer.com/content/pdf/10.1134/S002136401611014X.pdf
\begin{align}
	H(\bm{k})_+ &= +\hbar C(k_z - Q) - \hbar v \bm{\sigma}\cdot (\bm{k} - Q \hat{k}_z)\nonumber\\
	H(\bm{k})_- &= -\hbar C(k_z + Q) + \hbar v \bm{\sigma}\cdot (\bm{k} + Q \hat{k}_z)
	\label{zyuzin-model}
\end{align}
which describes two Weyl points with chirality $\pm 1$ separated by a distance $2Q$ in momentum space along the $\hat{k}_z$ direction. For the velocity units used in Table \ref{weyl-data}, $\hbar = 1$, but is included here for clarity. In this sign convection, positive $v$ describes a negatively charged Weyl point at $k=Q$, and a positive Weyl point at $k=-Q$. Changing the sign of $v$ interchanges the two Weyl points. The parameter $C$ controls the tilting of the Weyl cones, with positive $C$ describing a tilting of the Weyl cones \textit{inward} towards $k_z=0$, and negative $C$ corresponding to both cones tilting \textit{outward}, away from $k_z=0$:

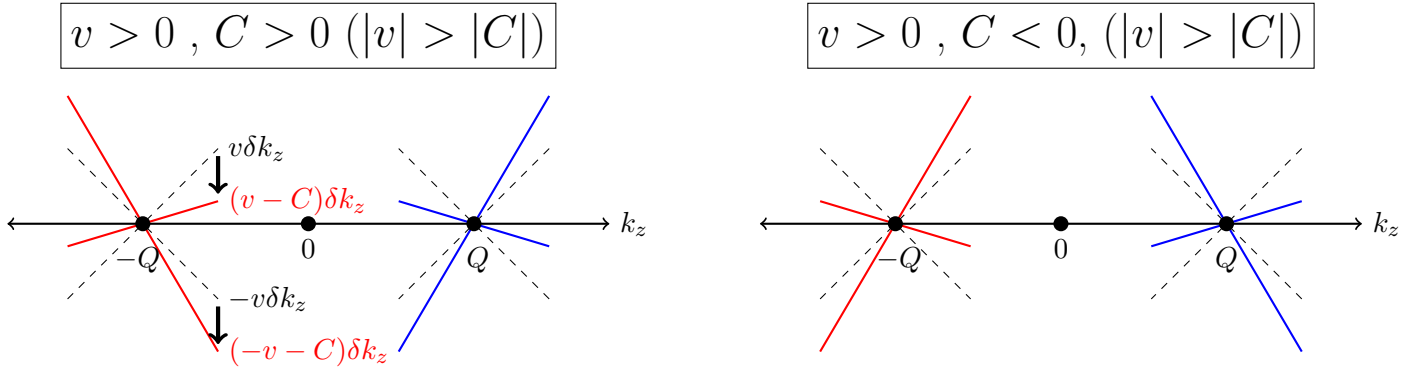
\begin{figure}[H]
\begin{center}
	\begin{tikzpicture}
	
	%Case 1
	\node[draw] at (0,2.5){\Large $v>0$ , $C>0$ ($|v|>|C|$)};
	\draw[black, thick,<->] (-4,0) -- (4,0) node[anchor=west]{$k_z$};
	\fill[black] (0,0) circle (0.1) node[anchor=north,yshift=-2pt]{0};
	
	%left Weyl
	\draw[black,dashed] (-3.2, 1)--(-1.2,-1)node[anchor=west]{$-v\delta k_z$}; 
	\draw[black,dashed] (-3.2,-1)--(-1.2, 1)node[anchor=west]{$v\delta k_z$};
	\draw[red,thick](-3.2, 1.7)--(-1.2,-1.7)node[anchor=west]{$(-v-C)\delta k_z$};
	\draw[red,thick](-3.2,-0.3)--(-1.2, 0.3)node[anchor=west]{$(v-C)\delta k_z$};
	\fill[black] (-2.2,0) circle (0.1) node[anchor=north, yshift=-5pt, xshift=-2pt]{\small $-Q$};
	
	\draw[black,ultra thick,->] (-1.2,0.9)--(-1.2,0.4);
	\draw[black,ultra thick,->] (-1.2,-1.1)--(-1.2,-1.6);
	
	%right Weyl
	\draw[black,dashed] (3.2,1)--(1.2,-1);% node[anchor=east]{$-v\delta k_z$}; 
	\draw[black,dashed] (3.2,-1)--(1.2,1);% node[anchor=east]{$v\delta k_z$};
	\draw[blue,thick] (3.2,1.7)--(1.2,-1.7);% node[anchor=east]{$(-v-C)\delta k_z$};
	\draw[blue,thick] (3.2,-0.3)--(1.2,0.3);% node[anchor=east]{$(v-C)\delta k_z$};
	\fill[black] (2.2,0) circle (0.1) node[anchor=north, yshift=-5pt, xshift=1.5pt]{\small $Q$};
	
	%separation
	%\draw[black,thick,|-|] (-3,1.3)--(3,1.3) node[midway,anchor=south]{\Large $2Q$};

	%Case 2
	\node[draw] at (10,2.5){\Large $v>0$ , $C<0$, ($|v|>|C|$)};
	\draw[black, thick,<->] (6,0) -- (14,0) node[anchor=west]{$k_z$};
	\fill[black] (10,0) circle (0.1) node[anchor=north,yshift=-2pt]{0};
	%left Weyl
	\draw[black,dashed] (6.8, 1)--(8.8,-1); 
	\draw[black,dashed] (6.8,-1)--(8.8, 1);
	\draw[red,thick](6.8,-1.7)--(8.8, 1.7);
	\draw[red,thick](6.8, 0.3)--(8.8, -0.3);
	\fill[black] (7.8,0) circle (0.1) node[anchor=north, yshift=-5pt, xshift=1.5pt]{\small $-Q$};
	%right Weyl
	\draw[black,dashed] (13.2, 1.0)--(11.2,-1.0);
	\draw[black,dashed] (13.2,-1.0)--(11.2, 1.0);
	\draw[blue,thick]   (13.2,-1.7)--(11.2, 1.7);
	\draw[blue,thick]   (13.2, 0.3)--(11.2,-0.3);
	\fill[black] (12.2,0) circle (0.1) node[anchor=north, yshift=-5pt]{\small $Q$};
	\end{tikzpicture}
\end{center}
\caption{\textbf{Type-I Weyl points.}}
\end{figure}
\vspace{-0.2cm}

In the figure above, the dashed lines, denote untilted Type-I cones (second terms in equation \ref{zyuzin-model}). Taking the $C$-dependent terms into account tilts the Weyl cones. By increasing $C$ until it exceeds the Fermi velocity $|C|>|v|$, the Weyl cones can be tilted below the horizontal, transitioning from Type-I to Type-II Weyl points:

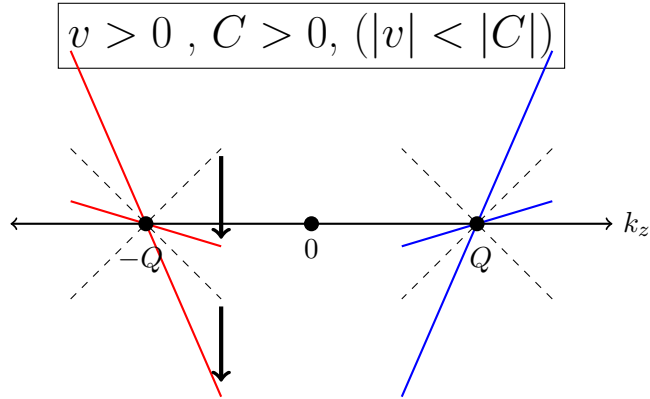
\begin{figure}[H]
\begin{center}
	\begin{tikzpicture}
	%Case 3 (Type-II)
	\node[draw] at (0,2.5){\Large $v>0$ , $C>0$, ($|v|<|C|$)};
	\draw[black, thick,<->] (-4,0) -- (4,0) node[anchor=west]{$k_z$};
	\fill[black] (0,0) circle (0.1) node[anchor=north,yshift=-2pt]{0};
	
	%left Weyl
	\draw[black,dashed] (-3.2, 1)--(-1.2,-1);%node[anchor=west]{$-v\delta k_z$}; 
	\draw[black,dashed] (-3.2,-1)--(-1.2, 1);%node[anchor=west]{$v\delta k_z$};
	\draw[red,thick](-3.2, 2.3)--(-1.2,-2.3);%node[anchor=west]{$(-v-C)\delta k_z$};
	\draw[red,thick](-3.2,0.3)--(-1.2, -0.3);%node[anchor=west]{$(v-C)\delta k_z$};
	\fill[black] (-2.2,0) circle (0.1) node[anchor=north, yshift=-5pt, xshift=-2pt]{\small $-Q$};
	
	\draw[black,ultra thick,->] (-1.2,0.9)--(-1.2,-0.2);
	\draw[black,ultra thick,->] (-1.2,-1.1)--(-1.2,-2.1);
	
	%right Weyl
	\draw[black,dashed] (3.2,1)--(1.2,-1);% node[anchor=east]{$-v\delta k_z$}; 
	\draw[black,dashed] (3.2,-1)--(1.2,1);% node[anchor=east]{$v\delta k_z$};
	\draw[blue,thick] (3.2,2.3)--(1.2,-2.3);% node[anchor=east]{$(-v-C)\delta k_z$};
	\draw[blue,thick] (3.2,0.3)--(1.2,-0.3);% node[anchor=east]{$(v-C)\delta k_z$};
	\fill[black] (2.2,0) circle (0.1) node[anchor=north, yshift=-5pt, xshift=1.5pt]{\small $Q$};

	\end{tikzpicture}
\end{center}
\caption{\textbf{Type-II Weyl points.}}
\end{figure}
\vspace{-0.2cm}

\begin{figure}[hb]
	\centering
	\begin{subfigure}[b]{0.32\textwidth}
		\centering
		\includegraphics[width=\textwidth]{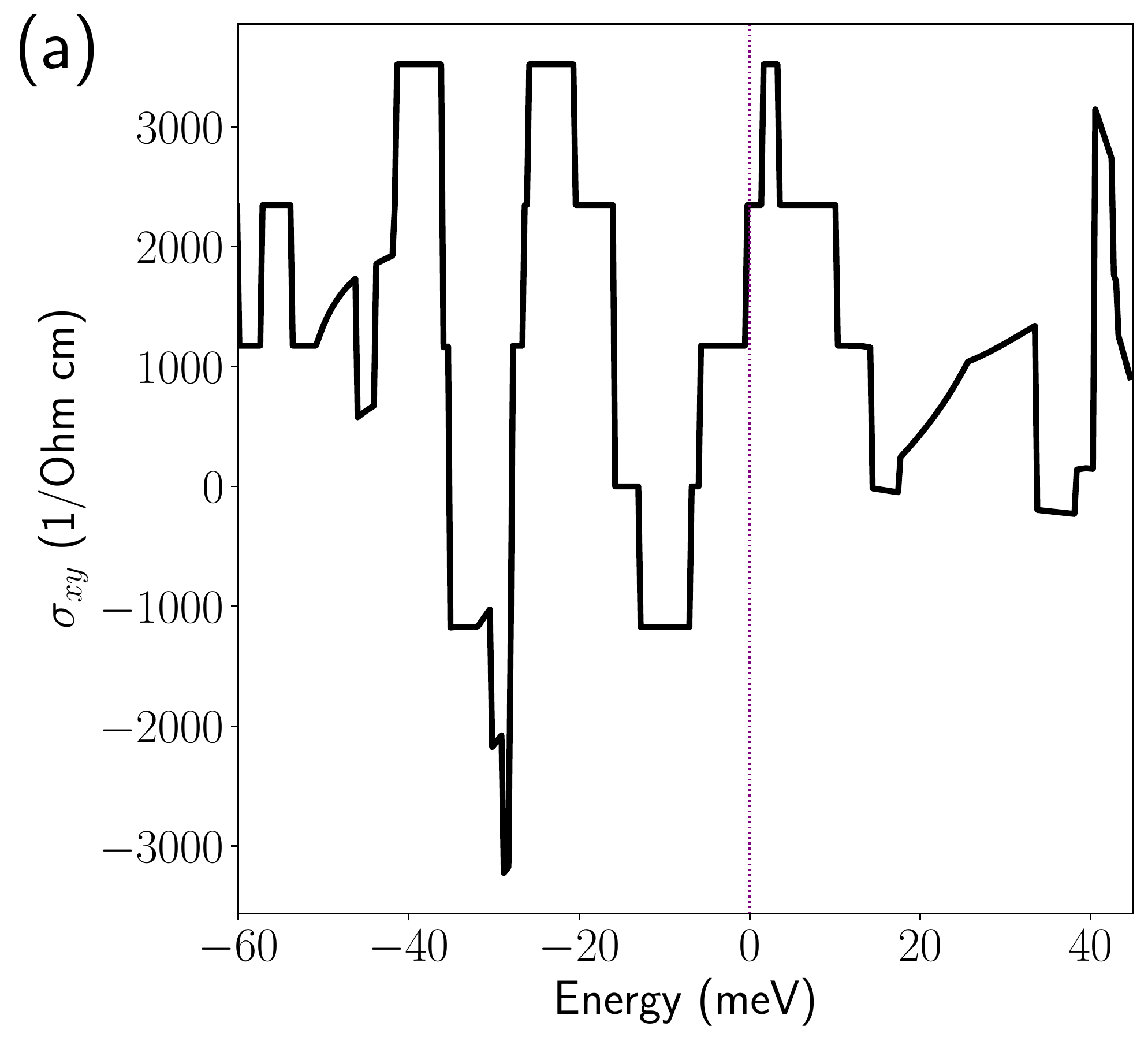}
	\end{subfigure}
	\hfill
	\begin{subfigure}[b]{0.32\textwidth}
		\centering
		\includegraphics[width=\textwidth]{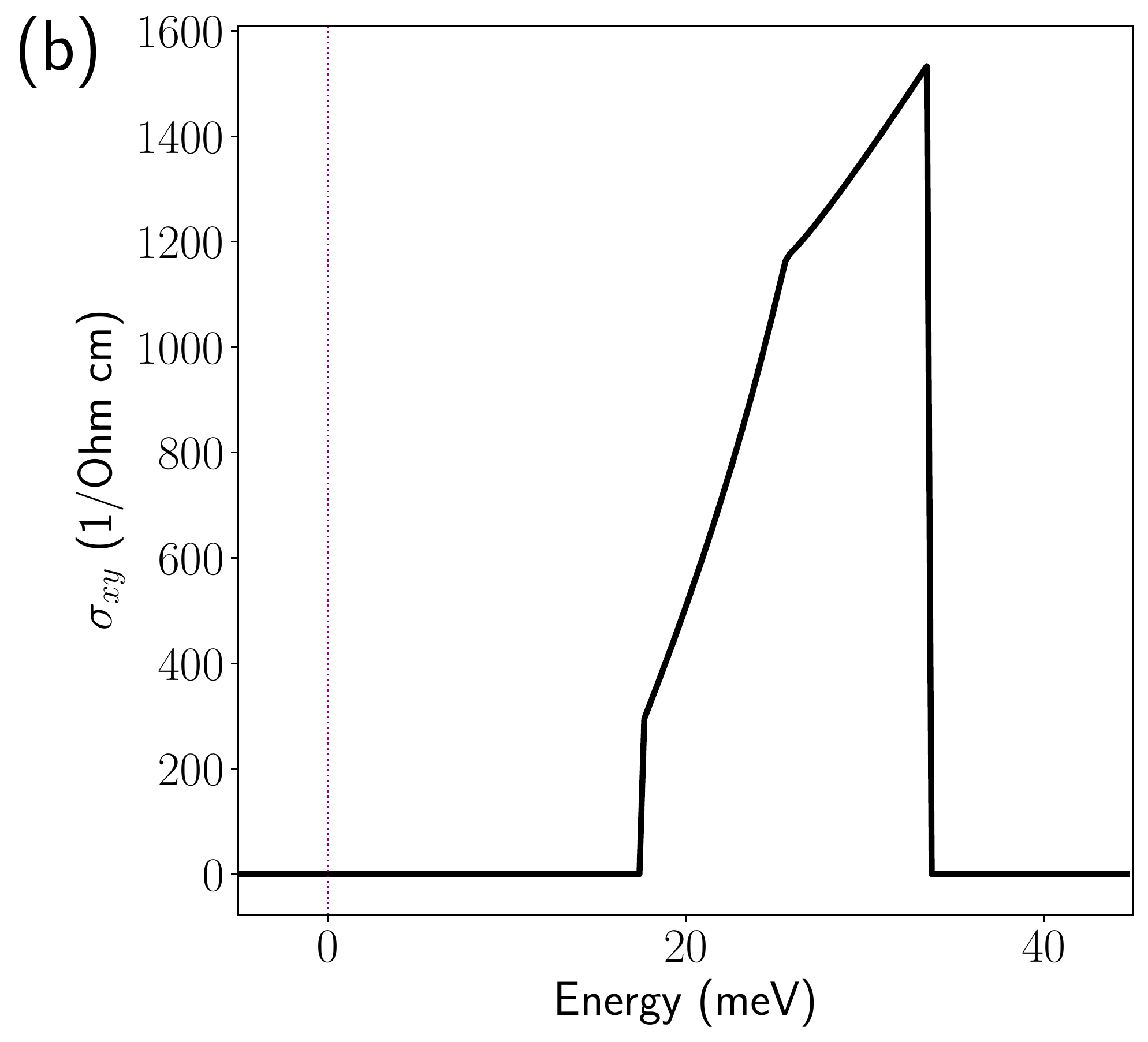}
	\end{subfigure}
	\hfill
	\begin{subfigure}[b]{0.32\textwidth}  
		\centering
		\includegraphics[width=\textwidth]{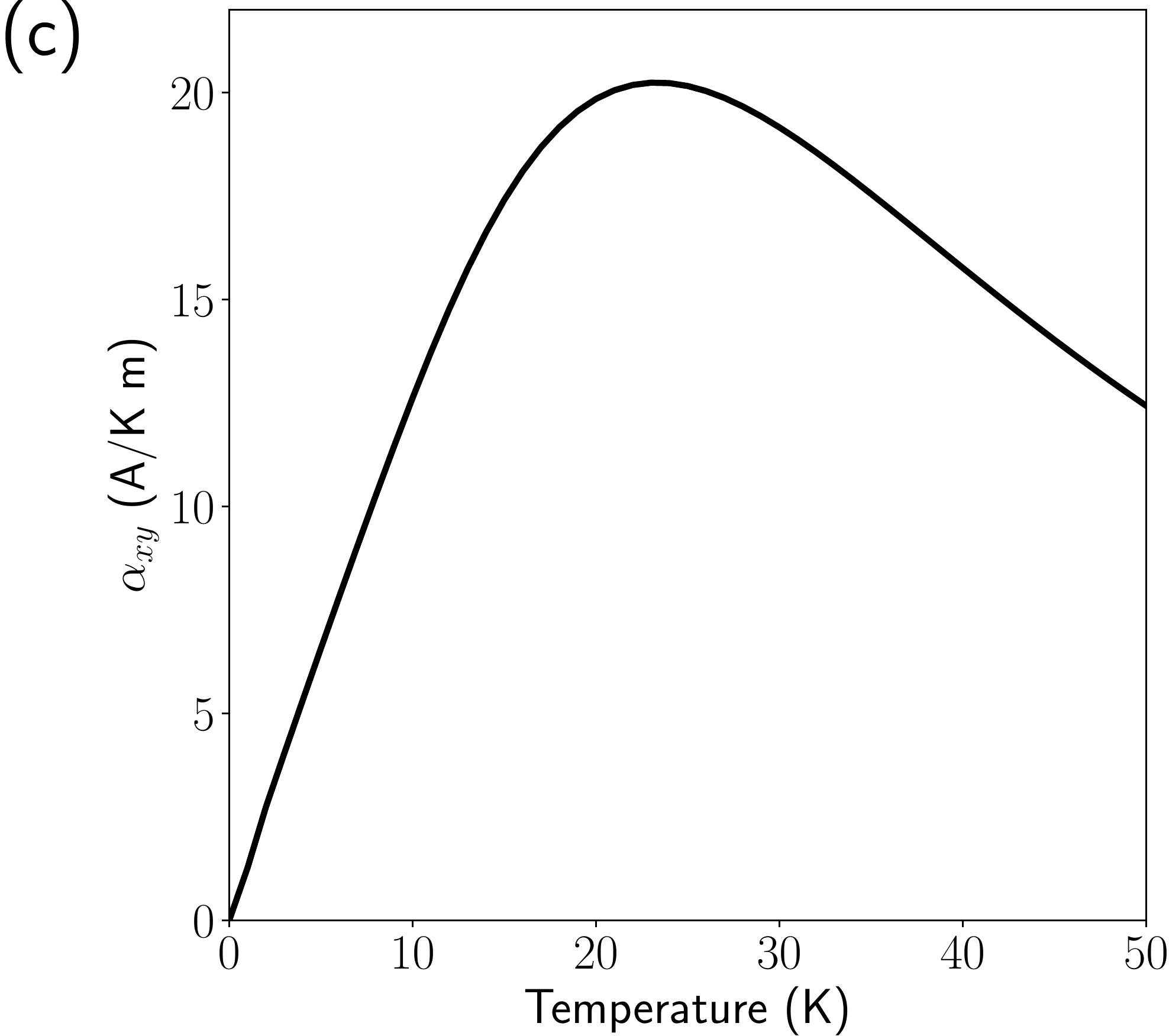} 
	\end{subfigure}
	\caption{\textbf{Calculated anomalous Hall effect and anomalous Nernst effects for UCo$_{0.8}$Ru$_{0.2}$Al.} a) Anomalous Hall effect computed for all Weyl points in UCo$_{0.8}$Ru$_{0.2}$Al. b) The anomalous Hall effect for set of Weyl points at +26 meV  c) The anomalous Nernst effect as a function of temperature computed with the chemical potential set at +26 meV.} 
	\label{ahe-ane}
\end{figure}

For this model, the anomalous Hall contribution in the limit of zero temperature, can be written:
\begin{align}
	\sigma_{xy} =- \frac{e^2}{8\pi^2} \int_{\Lambda-Q}^{-\Lambda-Q} dk_z \bigg[ \text{sign}&(k_z) \theta(v^2k_z^2-(Ck_z-(\epsilon-\epsilon_W))^2)  \nonumber\\
	&+\frac{vk_z}{|Ck_z-(\epsilon-\epsilon_W)|} (1-\theta(v^2k_z^2 - (Ck_z-(\epsilon-\epsilon_W))^2) )\bigg] \nonumber\\
	  + \frac{e^2}{8\pi^2} \int_{\Lambda+Q}^{-\Lambda+Q} dk_z \bigg[ \text{sign}&(k_z) \theta(v^2k_z^2-(-Ck_z-(\epsilon-\epsilon_W))^2) \nonumber\\
	 &+\frac{vk_z}{|-Ck_z-(\epsilon-\epsilon_W)|} (1-\theta(v^2k_z^2 - (-Ck_z-(\epsilon-\epsilon_W))^2) )\bigg] \nonumber\\
\end{align}
where $e_W$ is the energy at which the Weyl point is located, $\theta(x)$ is the Heaviside step function, and $\Lambda$ is an effective momentum cutoff. The first integral in the expression is the contribution from the Weyl at $k_z=Q$, and the second is from the Weyl at $k_z=-Q$. The terms in the first integral can be understood as a contribution of $v k_z/ |Ck_z-(\epsilon-\epsilon_W)|$ when $-\mu/(V-C)<k_z<mu/(V+C)$ and sign$(k_z)$ otherwise (flip signs for terms in the second integral).\\
\indent In turn, the anomalous Nernst effect can be computed from the zero-temperature anomalous Hall through the following formula \cite{xiao2006berry}:
\begin{align}
\alpha(T,\mu) &= -\frac{1}{e} \int d\epsilon \left( \frac{\partial f_{\text{FD}} }{\partial \mu} \right) \sigma(0,\epsilon) \frac{\epsilon-\mu}{T} 
= -\frac{1}{e} \int d\epsilon \frac{e^{(\epsilon-\mu)/(k_BT)}}{k_BT \left( e^{(\epsilon-\mu)/(k_BT)} +1 \right)^2} \sigma(0,\epsilon) \frac{\epsilon-\mu}{T} \nonumber\\
&= -\frac{1}{eT} \int  w(\frac{\epsilon-\mu}{k_BT}) \sigma(0,\epsilon) d\epsilon
\end{align}
where $ f_{\text{FD}}$ is the Fermi-Dirac distribution, and $w(x)= xe^x/(e^x+1)^2$ is a weight function introduced to simplify the expression.\\
\indent The computed anomalous Hall effect for the Weyl points listed in Table \ref{weyl-data} is shown in Figure \ref{ahe-ane}a. A large value of $\sigma_{xy}$ appears just above the Fermi level, suggesting that a collective contribution from several Weyl points could explain the large experimentally observed values. This plot must be interpreted very carefully for a number of reasons. Firstly, the anomalous Hall contribution for each pair of Weyl points depends on their $k_z$ separation, which in turn depends on the uranium magnetic moments. As we have previously mentioned, these are difficult to reproduce numerically. Secondly, while we take a relatively dense initial grid of $30 \times 30 \times 30$ k-points to search for topological features, the dense distribution of the Weyl points we found suggests the possibility that some initial $k$-cubes may contain multiple Weyl points, meaning there may be yet more additional Weyl points missed by our procedure. Thirdly, this calculation does not include contributions from nodal lines and other sources. Finally, UCo$_{1-x}$Ru$_x$Al is magnetic for dopings $x = 0.005 - 0.78$, so outside of this range, our model would not apply.\\
\indent We also draw attention to the anomalous Hall effect contribution coming from the Weyl point located at (0.0, 0.51,-0.02) and 26 meV above the Fermi energy (and its symmetry related partners) (Fig~\ref{ahe-ane}b),c)). The anomalous Nernst contribution of these points is computed by setting the chemical potential at +26 meV and evaluating the above integral.
This set of Weyl points is close to a Lifshitz transition between Type-I and Type-II tilting.  For tilted Weyl points near the Lifshitz transition, ${\partial \sigma_{xy}}/{\partial \epsilon}$ diverges logarithmically at the critical tilt, resulting in large anomalous Hall and Nernst effects, as well as an amplification of $\alpha_{xy}$ relative to $\sigma_{xy}$ \cite{sakai2018giant}. In UCoAl, these Weyl points are located +26 meV above the Fermi energy, corresponding to a doping of $x=0.07$, and are sufficient to explain the large anomalous Hall ($\sim 1200$ [Ohm cm]$^{-1}$), large anomalous Nernst ($\sim 20$ A/K m), and a $\alpha_{xy}/\sigma_{xy}$ ratio which exceeds $k_B/e$ observed in this material.\\
\indent We reiterate that the Weyl points in this material, along with their associated Berry curvature and other observable properties, are sensitive to the uranium magnetic moments and additional electronic renormalization effects. This means that in the undoped case, the absence of magnetism would guarantee a zero anomalous Hall effect, even though Fig~\ref{ahe-ane}a suggests a large value. Additionally, as it is difficult to exactly reproduce the uranium magnetic moment $\mu_U$, the critical set of Weyl points identified above may be located at a lower energy corresponding to the $x=0.2$ doping level.\\
\indent A final consideration is that the magnetism in UCo$_{1-x}$Ru$_{x}$Al disappears at higher temperatures, meaning the anomalous Nernst effect will rapidly go to zero as it approaches the magnetic transition.\\
\indent To summarize, our calculations yield a large number of topological features, including Weyl points, triple points, and nodal lines. Some Weyl nodes are sufficient to explain the the large anomalous Hall and anomalous Nernst observed in UCo$_{0.8}$Ru$_{0.2}$Al. Additional calculations and measurements are needed to identify which features are most responsible.

\section{Measurement setup}

\begin{figure}[h]
	\includegraphics[width=0.7\linewidth]{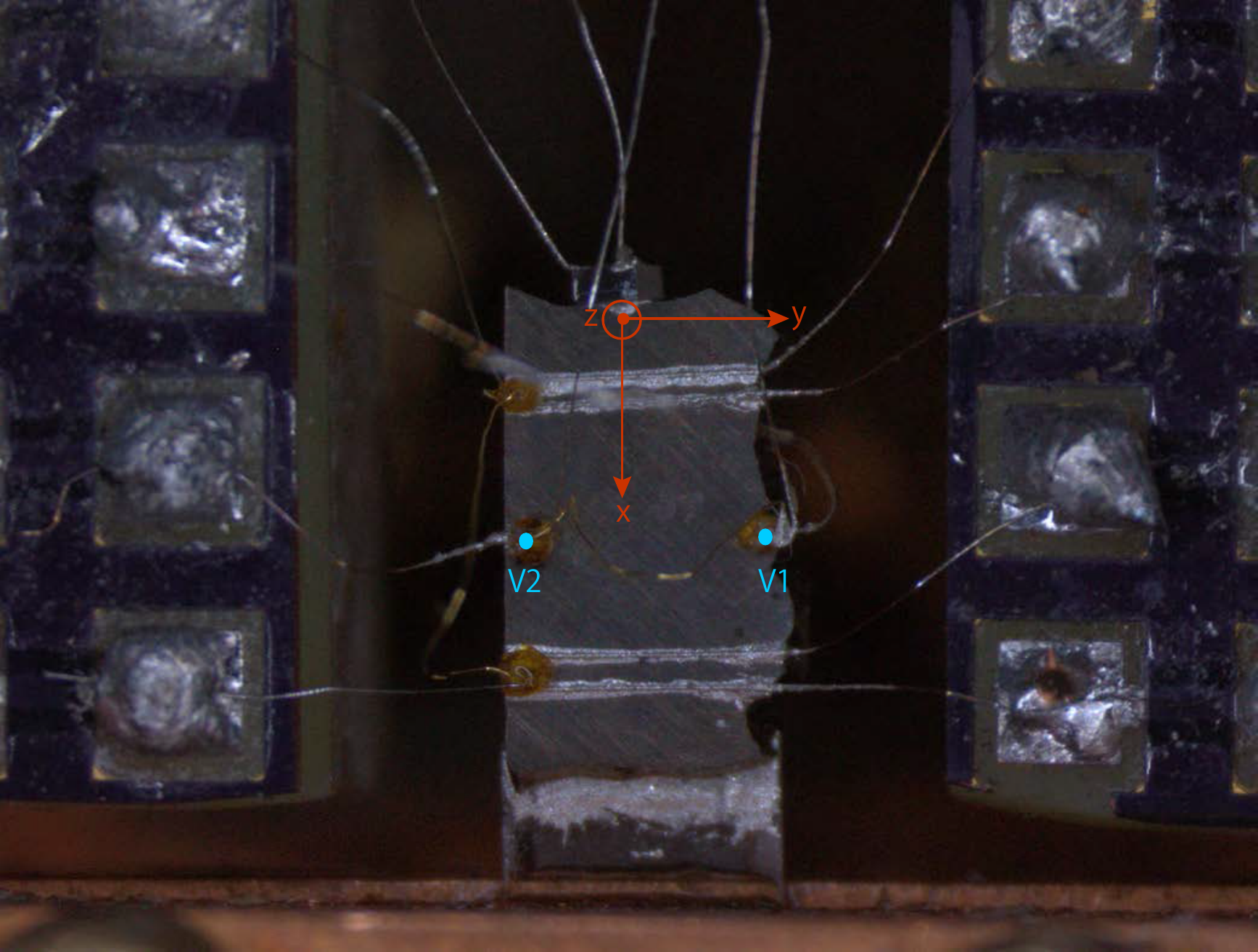}
	\caption{\label{setup} 
		{\textbf{Photograph of our experimental setup.} The magnetic field is applied parallel to $c$-axis. Photo credit: T. Asaba (Los Alamos National Laboratory).} 
	}
\end{figure}

Fig. \ref{setup} shows the transport measurement setup. A resistor is attached to the sample as a heater, generating the heat flow. The other end is glued to the sapphire substrate via silver epoxy. The low-thermal conductivity phosphor-bronze copper wires were attached to the sample and heater to minimize the heat dissipation through the electrical leads. Home-made gold-iron and chromel thermocouples were calibrated and used to measure the temperature gradient in the linear response regime. We define our coordinates as shown in red in Fig. \ref{setup}. The heat and current flow is along the $x$ direction, which is also parallel to the $a$-axis of the crystal. $y$ and $z$ are parallel to the $a^*$- and the $c$-axis of the crystal, respectively. 

\begin{figure}[h]
	\includegraphics[width=0.5\linewidth]{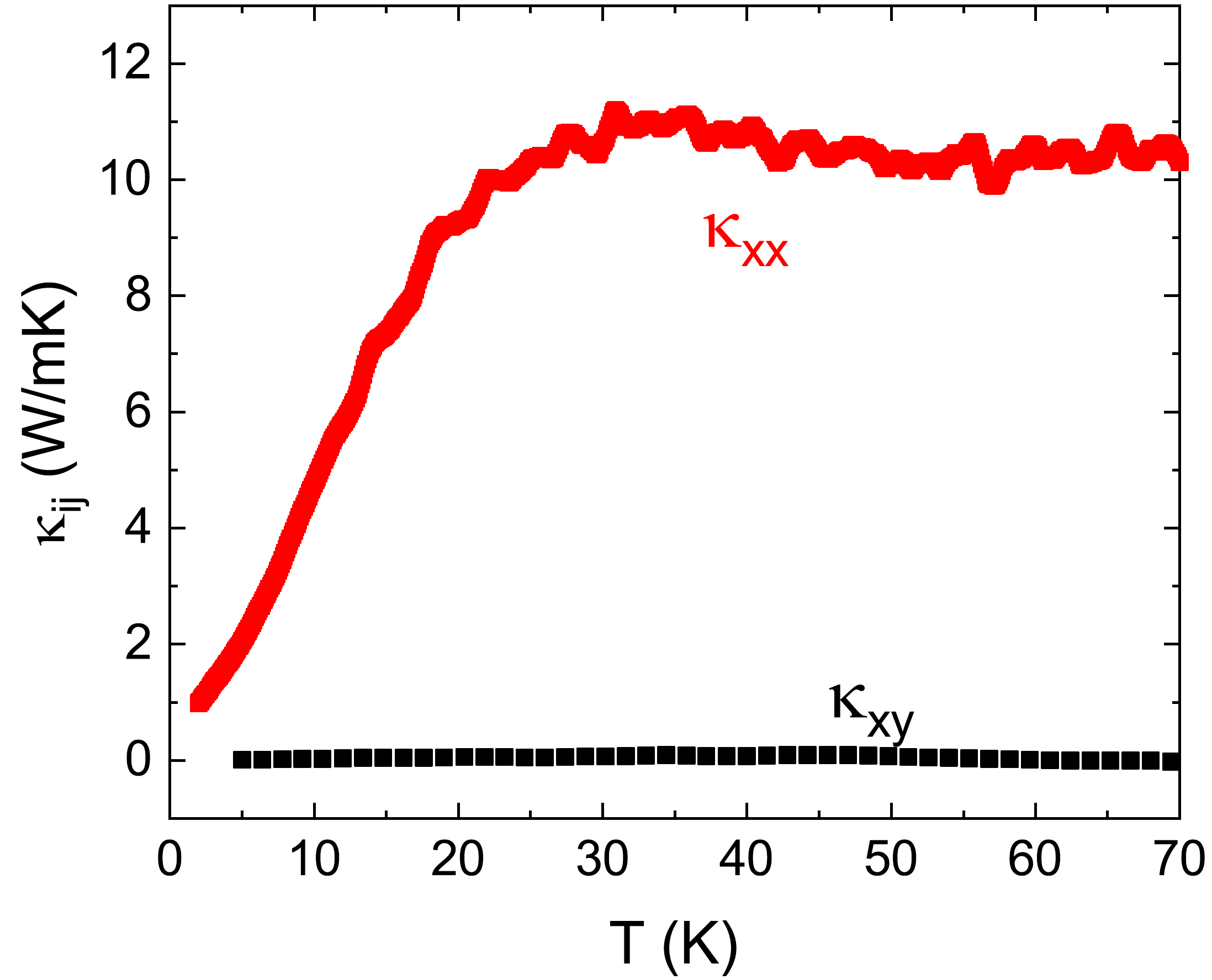}
	\caption{\label{thermalconductivity} 
		{\textbf{Thermal conductivity of UCo$_{0.8}$Ru$_{0.2}$Al.} Longitudinal (red) and transverse (black) thermal conductivity as a function of temperature.} 
	}
\end{figure}

The electric and heat currents follow the equations below.
\begin{equation}\label{electronEQ}
J_e = \boldsymbol{\sigma} E - \boldsymbol{\alpha} \nabla T
\end{equation}
\begin{equation}\label{heatEQ}
J_q = T \boldsymbol{\alpha} E - \boldsymbol{\kappa} \nabla T
\end{equation}
where $J_e$ and $J_q$ are electron and heat flow, $\boldsymbol{\sigma}$, $\boldsymbol{\alpha}$ and $\boldsymbol{\kappa}$ are conductivity, thermoelectric (Peltier) and thermal conductivity tensors, and $E$ is the electric field, respectively. In Eq. \ref{heatEQ}, the first term on the right hand side is negligible compared to the second term for our sample. Therefore, the expressions for the thermal conductivity are $\kappa_{xx}$=$-J_q \nabla_x T$/(($\nabla_x T)^2+(\nabla_y T)^2$) and $\kappa_{xy}$=$-J_q\nabla_y T$/(($\nabla_x T)^2+(\nabla_y T)^2$) where $\nabla_y T$ = $+\Delta T_y/w$ with $\Delta T_y$ = ($T_1$ - $T_2$) and   $\nabla_x T$ = $-\Delta T_x$/w with $\Delta T_x$ = ($T_{hot} - T_{cold}$) being the temperature drop along the length of the sample in the $x$ direction. As shown in Fig. \ref{thermalconductivity}, we find $\kappa_{xy} \ll \kappa_{xx}$ and therefore $\Delta T_y \ll \Delta T_x$. In this setup the electrical current is zero (ie. $J_e$ = 0). By defining the thermoelectric tensor $\boldsymbol{S} \equiv \boldsymbol{\rho\alpha}$ where $\rho$ is resistivity tensor, we obtain
\begin{equation}
E_x=S_{xx}\nabla_xT+S_{xy}\nabla_yT
\end{equation}
\begin{equation}
E_y=-S_{xy}\nabla_xT+S_{yy}\nabla_yT.
\end{equation}

Given that $\Delta T_y \ll \Delta T_x$ and $S_{xx} \approx S_{yy}$, these equations further reduce to $S_{xx}=-E_xl/\Delta T_x$ and $S_{xy}= E_yl/\Delta T_x$, which is our definition of the Seebeck and Nernst effects.

In our coordinate, the Nernst effect takes the modern convention and is defined as S$_{xy}$ as $-(V1-V2)l/\Delta T_x$w, where the Hall effect is defined as $\rho_{xy}$= $(V1-V2)t/I_x$. Here $V1$ and $V2$ are the voltages measured at the points shown in cyan in the figure, $I_x$ is the current along the $x$ direction. $l$, $w$, and $t$ are the length, width and thickness of the sample, respectively. In both cases, the positive field is applied along the $z$ direction.
%%%%%%%%%Figure1$$$$$$$$$$$
%%%%%%%%%Figure1$$$$$$$$$$$

%%%%%%%%%Figure1$$$$$$$$$$$
%%%%%%%%%Figure1$$$$$$$$$$$

\section{Two components of the Peltier coefficient $\alpha$}
\begin{figure}[h]
	\includegraphics[width=0.5\linewidth]{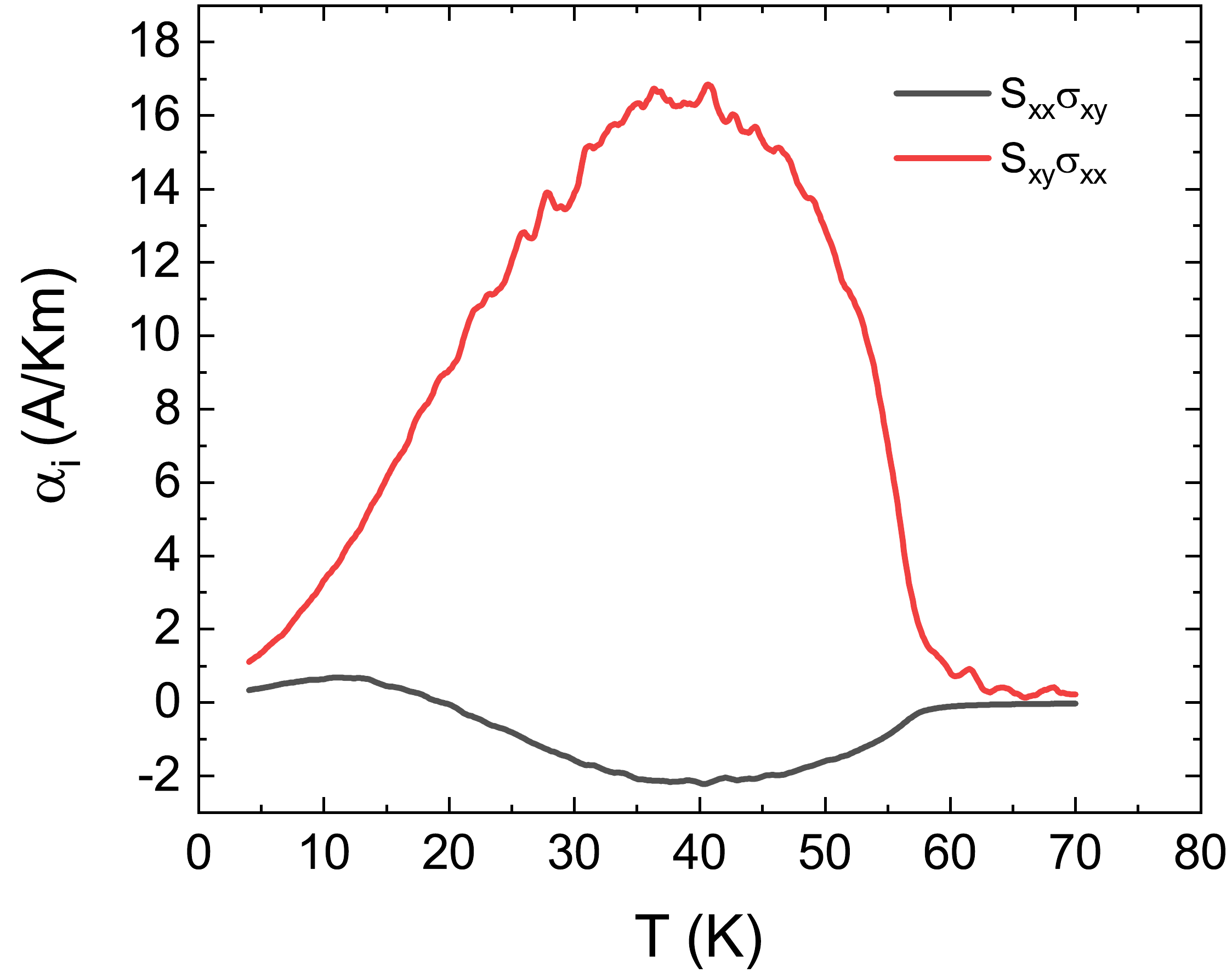}
	\caption{\label{twoalpha} 
		\textbf{Two Peltier components.} Two components of $\alpha_{xy}$=$\alpha_1$+$\alpha_2$=$S_{xx}\sigma_{xy}$+$S_{xy}\sigma_{xx}$ are shown as a function of temperature.
	}
\end{figure}

The Peltier coefficient is the sum of two terms $\alpha_{xy} = \alpha_1 + \alpha_2$ with $\alpha_1 = S_{xx}\sigma_{xy}$ and $\alpha_2 = S_{xy}\sigma_{xx}$. Fig. \ref{twoalpha} shows the temperature dependence of $\alpha_1$ and $\alpha_2$. The latter component is almost one order of magnitude larger than the former one at $T$ = 41 K, dominating the Peltier coefficient. This is one of the significant features of UCo$_{0.8}$Ru$_{0.2}$Al, as most known materials are dominated by $S_{xx}\sigma_{xy}$.

%%%%%%%%%Figure1$$$$$$$$$$$
%%%%%%%%%Figure1$$$$$$$$$$$

\section{Specific heat measurements}
\begin{figure}[h]
	\includegraphics[width=0.5\linewidth]{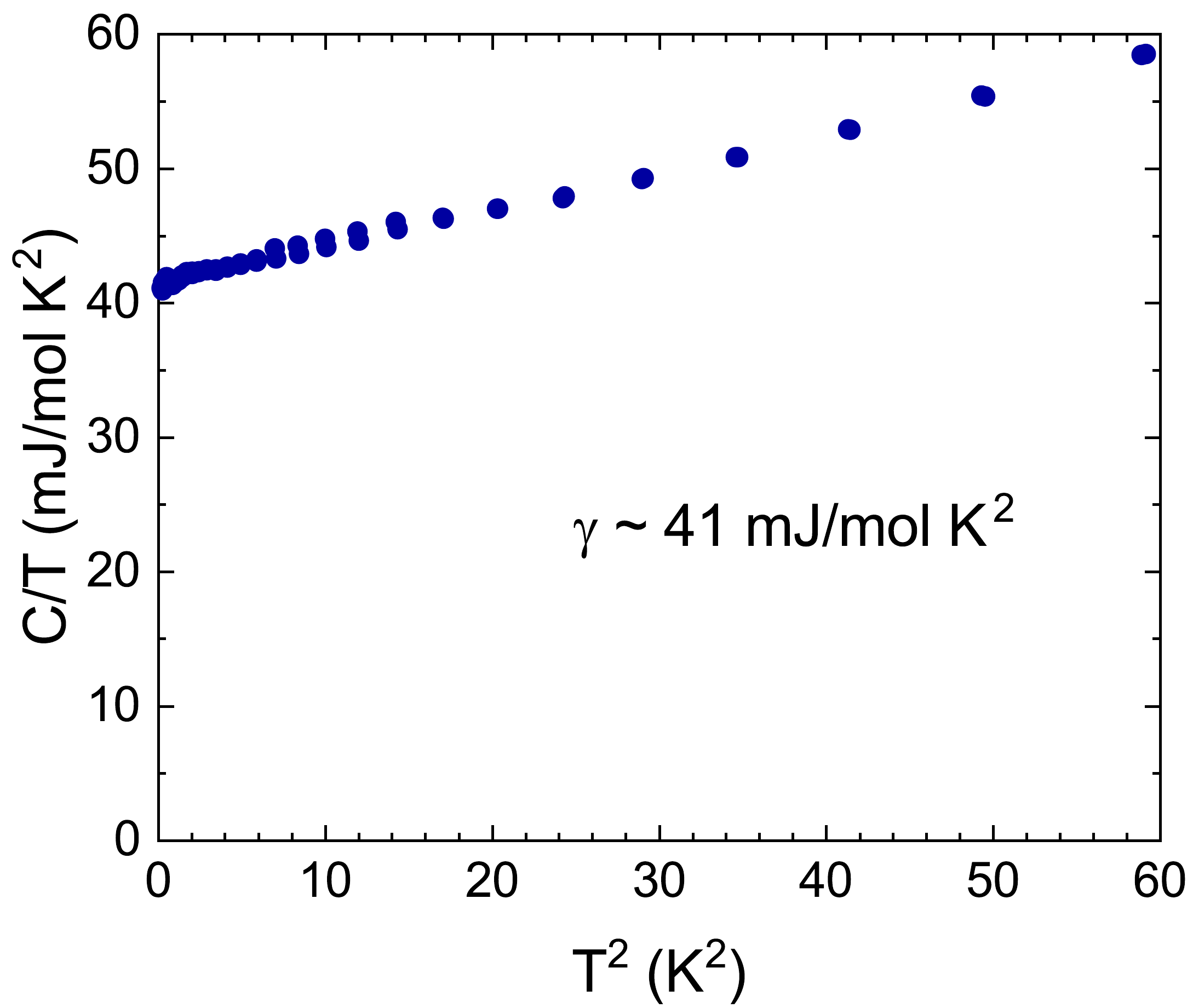}
	\caption{\label{HC} 
	    \textbf{ The $C/T$ as a function of temperature squared $T^2$.} The extrapolated value of $C/T$ to $T$=0 gives the Sommerfeld coefficient $\gamma$=41 mJ/mol K$^2$.}
\end{figure}

The low-temperature specific heat measurement data is shown in Fig. \ref{HC}. The specific heat over temperature is plotted as a function of temperature squared. The Sommerfeld coefficient $\gamma$=41 mJ/mol-U K$^2$ indicates that the system is moderately correlated. The Fermi temperature, $T_F = E_F/k_B$ is estimated by the equation \cite{behnia2004thermoelectricity}
\begin{equation}
\gamma = \frac{\pi^2 k_B^2 n}{2E_F}
\end{equation}
where $\gamma$ is the Sommerfeld coefficient in units of J/K$^2$m$^3$, $k_B$ is Boltzmann's constant and $n$ is the carrier density. Here, we assume that there is one carrier per unit cell. The estimated Fermi temperature is 930 K. 
The density of states computed from our LSDA calculations is shown in Fig. 1 of the main text. From this the computed Sommerfeld coefficient can be obtained $\gamma_{band}$ = $(\pi^2 k_B^2/3)N_\text{LSDA}(0)$ = 13 mJ/mol-U K$^2$. A comparison with the experimental $\gamma$ gives a mass enhancement of $\sim$ 3.

\begin{figure}[h]
	\includegraphics[width=0.5\linewidth]{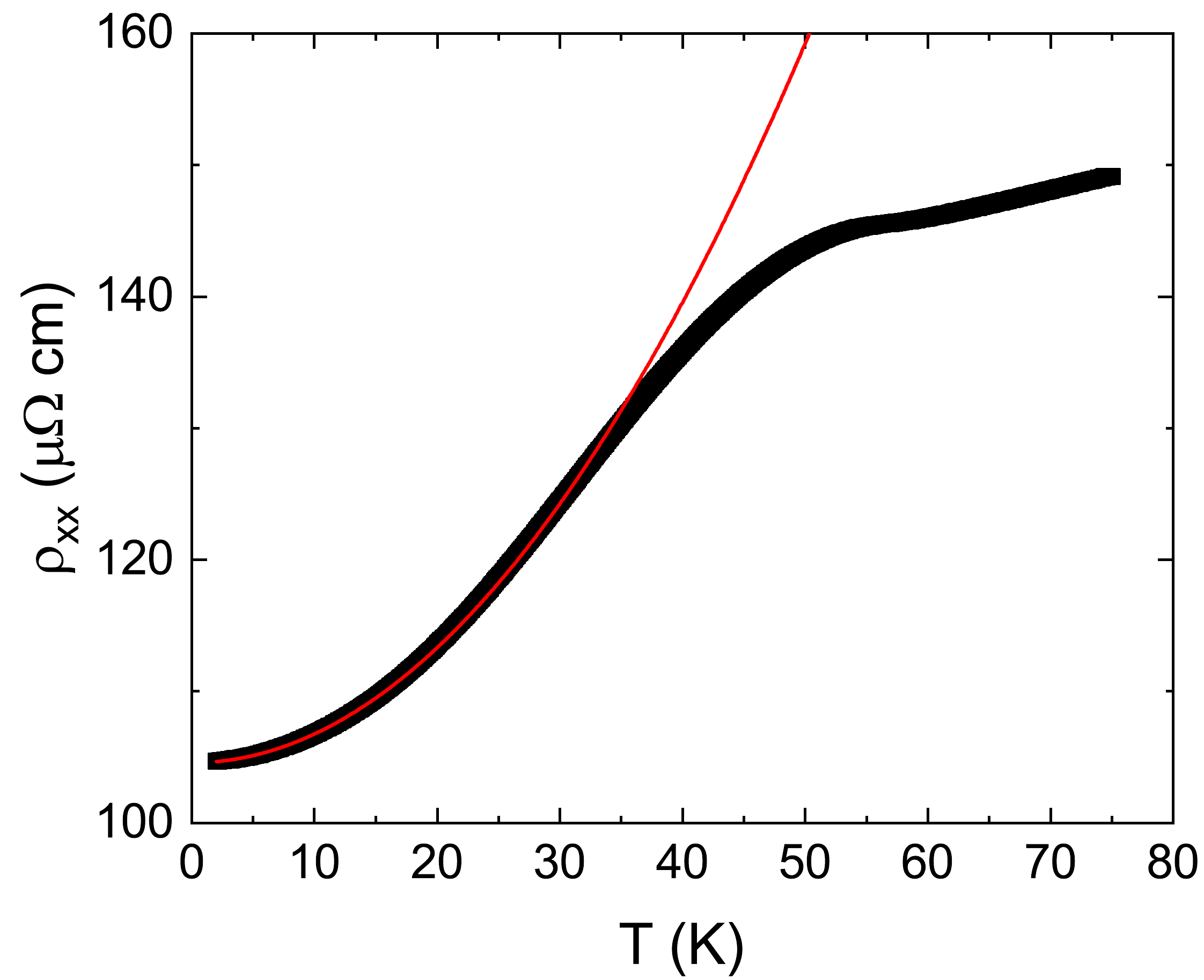}
	\caption{\label{kadowaki} 
		{\textbf{The low temperature resistivity vs temperature plot.} The red line is a quadratic fit to $\rho$= $\rho_0$+$AT^2$ up to 15 K resulting in $A$ = 2.26 $\times$10$^{-8}\Omega$cm/K$^2$.} 
	}
\end{figure}

We also calculated the Kadowaki-Woods ratio $A/\gamma^2$, where $A$ is the $T^2$ coefficient of resistivity. $A$ is obtained by the quadratic fitting of resistivity as shown in Fig. \ref{kadowaki}. The obtained value of $A$ is 2.26 $\times$ 10$^{-8}$ $\Omega$cm/K$^2$. Accordingly, $A/\gamma^2$ of UCo$_{0.8}$Ru$_{0.2}$Al is estimated to be 1.3 $\times$ 10$^{-5}$ ($\mu\Omega$cm/K$^2$)/(mJ/mol K$^2$)$^2$, which is close to the universal value of $\sim$ 1$\times$ 10$^{-5}$ ($\mu\Omega$cm/K$^2$)/(mJ/mol K$^2$)$^2$ \cite{kadowaki1986universal}.
%%%%%%%%%Figure1$$$$$$$$$$$
%%%%%%%%%Figure1$$$$$$$$$$$

%%%%%%%%%Figure1$$$$$$$$$$$
%%%%%%%%%Figure1$$$$$$$$$$$

\section{Seebeck effect at low temperatures}
Similarly, the Fermi temperature is estimated from $S/T$ at the zero temperature limit through the equation \cite{behnia2004thermoelectricity}
\begin{equation}
S/T = \frac{\pi^2 k_B}{2e T_F}
\end{equation}
where $e$ is the charge of an electron. $S/T$ in the $T$=0 limit is 0.9 $\mu$V/K$^2$ giving an estimated Fermi temperature of 470 K.
Also, the positive thermopower at zero temperature limit indicates that hole-like carriers dominate the thermoelectric transport at low temperatures.